\documentclass[aps,twocolumn,preprintnumbers,superscriptaddress,amsmath,amssymb]{revtex4-2}
\usepackage{float}
\usepackage{graphicx}
\usepackage{dcolumn}
\usepackage{bm}
\usepackage{epstopdf}
\usepackage{nameref}
\usepackage{xcolor}
\usepackage{textcmds}
\usepackage{textcomp}
\usepackage{soul}
\usepackage[utf8]{inputenc}
\usepackage{siunitx}
\usepackage{csquotes}
\usepackage{amsbsy}
\usepackage{mathrsfs} 
\usepackage{amsmath}
\usepackage{bigints} 
\usepackage{amsthm,amssymb} 
\usepackage[english]{babel}
\usepackage[shortlabels]{enumitem}
\usepackage{float}
\usepackage{mathrsfs,bigints,mathtools,dsfont}
\usepackage[colorlinks=true,linkcolor=blue,citecolor=blue]{hyperref}%
\usepackage[toc]{appendix}
\usepackage[normalem]{ulem}

\hypersetup{
  colorlinks   = true,	
  urlcolor     = blue, 
  linkcolor    = blue, 
  citecolor   = blue 
}

\begin{document}

\title{Diffusion-induced instabilities promote cooperation in eco-evolutionary networks}

\author{Sourav Roy} \thanks{These authors contributed equally to this work.}
\email{sourav.roy738@gmail.com}
 \affiliation{Bagchi School of Public Health, Ahmedabad University,  Ahmedabad 380009, Gujarat, India}
\author{Md Sayeed Anwar} \thanks{These authors contributed equally to this work.}
\email{sayeedanwar447@gmail.com}
\affiliation{Department of Physical Sciences, Indian Institute of Science Education and Research Kolkata, Mohanpur 741246, Nadia, West Bengal, India}
\author{Timoteo Carletti}
\email{timoteo.carletti@unamur.be}
\affiliation{Department of Mathematics and Namur Institute for Complex Systems - naXys, University of Namur, rue Grafé 2, B 5000 Namur, Belgium}
\author{Matja{\v z} Perc}\email{matjaz.perc@gmail.com}
\affiliation{Faculty of Natural Sciences and Mathematics, University of Maribor, Koro{\v s}ka cesta 160, 2000 Maribor, Slovenia}
\affiliation{Community Healthcare Center Dr. Adolf Drolc Maribor, Ulica talcev 9, 2000 Maribor, Slovenia}
\affiliation{Department of Physics, Kyung Hee University, 26 Kyungheedae-ro, Dongdaemun-gu, Seoul 02447, Republic of Korea}
\affiliation{University College, Korea University, 145 Anam-ro, Seongbuk-gu, Seoul 02841, Republic of Korea}
\author{Dibakar Ghosh}
\email{dibakar@isical.ac.in}
\affiliation{Physics and Applied Mathematics Unit, Indian Statistical Institute, 203 B. T. Road, Kolkata 700108, India}




\begin{abstract}

\textcolor{black}{Understanding how cooperation persists despite the advantage of selfish behavior remains a central challenge in evolutionary dynamics. Classical models of public goods dilemmas predict dominance of defectors, yet natural and social systems often sustain cooperation. We study an eco-evolutionary public goods game on complex networks where cooperators and defectors diffuse at different rates. When the isolated system is in a defector-dominated coexistence regime, faster dispersal of defectors than cooperators leads to a symmetry-breaking transition that produces localized clusters of cooperators. In heterogeneous networks, nodes with higher connectivity become significantly more likely to exhibit cooperative dominance. {\color{black} A degree-based mean-field reduction supports this result by showing that network connectivity controls an effective coupling strength proportional to node degree, thereby producing a bifurcation that separates defector-dominated and cooperative states.} We also address why not all hubs become cooperative by means of a multistability analysis. These results reveal how asymmetric mobility and heterogeneous connectivity jointly promote cooperation in structured populations.}
\end{abstract}

\maketitle

\section{Introduction} \label{intro}

The persistence of cooperation in social and biological systems has long been a central point of interest in evolutionary theory. This enduring inquiry traces back to foundational questions about how cooperative traits can survive natural selection when selfish behavior seems evolutionarily advantageous. In well-mixed populations, where each agent interacts with every other one without spatial constraints, traditional game-theoretic models, especially the public goods game, have predicted a grim outcome for cooperation \cite{szolnoki2010reward, balduzzi2018mechanics}. Within these models, defectors, by reaping benefits from shared resources without contributing, consistently outcompete cooperators. This exploitation-driven advantage leads to the eventual extinction of cooperative strategies \cite{nowak2006five, szabo2002phase}, by resulting in the breakdown of collective action and the depletion of essential environmental resources \cite{hardin2018tragedy, ostrom2008tragedy, sohel2024hypochaos}. Such theoretical outcomes underscore a central dilemma: if defection is individually advantageous, why is cooperation so pervasive in nature?

Contrary to these theoretical expectations, empirical observations reveal that cooperation is not a rare anomaly but a recurring feature of many biological and social systems, from microbial biofilms to human societies \cite{nowak2006evolutionary, nadell2016spatial, cremer2019chemotaxis, rand2014social}. These findings suggest that the assumptions of well-mixed populations are often too restrictive to capture real-world dynamics. Natural and social systems typically exhibit spatial structure, ecological feedback, and heterogeneous interactions \cite{roy2023time, betz2024evolutionary, chowdhury2021eco}. Incorporating such factors into evolutionary models has therefore become essential for understanding how cooperation can emerge and persist.

\textcolor{black}{Several mechanisms have been proposed to support cooperation in structured populations. Spatial structure and limited dispersal allow cooperators to cluster together, thereby protecting themselves from exploitation \cite{wakano2009spatial, wakano2011pattern}. Reaction–diffusion models have been particularly useful for studying such dynamics, as they describe how local interactions combined with dispersal can generate spatial patterns and cooperative clusters \cite{hernandez2021environmental}. However, these approaches typically assume continuous spatial domains, whereas many biological and social systems operate in discrete environments.}
\textcolor{black}{Microbial colonies often grow on patchy substrates, and human interactions frequently occur through discrete social or technological networks \cite{pan2023heterogeneous, pires2024rules}.}

These observations call for modeling frameworks capable of capturing discrete and heterogeneous interaction structures.

Complex networks provide a natural representation of such systems. Prominent examples include scale-free networks with highly connected hubs \cite{barabasi2003scale}, small-world networks characterized by short path lengths \cite{watts1998collective}, and modular networks with clustered connectivity \cite{andreas2016neural}. In these systems, interactions are governed not by physical distance but by network topology \cite{santos2005scale}. The Laplacian coupling formalism offers a convenient mathematical framework for modeling movement or diffusion across such networks, capturing how the state of each node relaxes toward that of its neighbors. This formulation has been widely used to study diffusion-driven instabilities and pattern formation in networked dynamical systems~\cite{nakao2010turing, gao2023turing}.

\textcolor{black}{Alternative mechanisms for promoting cooperation have also been explored. In particular, models of the iterated Prisoner’s Dilemma with partner choice or refusal have shown that allowing individuals to select or avoid interaction partners can stabilize cooperation by enabling cooperators to assort and avoid exploitation \cite{stanley1995iterated,ashlock1996preferential}. In such frameworks, cooperation arises through strategic interaction choices. In contrast, the mechanism considered here does not rely on partner selection. Instead, we focus on how mobility and spatial coupling alone, without active partner choice, can reshape eco-evolutionary dynamics in structured populations.}

Despite substantial progress in modeling cooperation on networks, several aspects of eco-evolutionary dynamics remain insufficiently explored. In particular, the combined effects of local group interactions, ecological feedback, and asymmetric mobility between strategies have rarely been integrated into a unified framework. In many real-world systems, cooperators and defectors differ in their mobility or dispersal rates. For instance, defectors may spread more rapidly across environments while cooperators remain localized within productive regions \cite{hu2024collective, zhang2025evolution, cheng2024evolution}. Such differences can generate spatial instabilities that reorganize population structure and potentially promote cooperation. From a dynamical perspective, these processes can be interpreted as diffusion-driven instabilities reminiscent of Turing pattern formation \cite{Turing1952}.

Motivated by these observations, we develop an eco-evolutionary public goods game model on complex networks that incorporates asymmetric mobility between cooperators and defectors through Laplacian-based diffusion \cite{gokhale2010evolutionary, gokhale2016eco, merris1994laplacian}. The model combines localized group interactions, ecological failure of public goods, and network-structured dispersal. Using analytical tools including dispersion relations and amplitude analysis \cite{nakao2010turing, AsllaniPRE2014, MuoloPRSA2024}, we identify conditions under which homogeneous coexistence becomes unstable and heterogeneous spatial patterns emerge. In this regime, cooperation becomes locally dominant even though the overall system favors defection. Our results further reveal that node connectivity plays a central role in shaping these dynamics, providing a mechanistic explanation for the emergence of cooperative hubs in structured populations. \textcolor{black}{A degree-based mean-field reduction reveals that node connectivity determines an effective coupling strength governing the local dynamics, explaining why highly connected nodes are statistically more likely to sustain cooperative dominance. Together, these findings demonstrate how the interplay between asymmetric mobility and heterogeneous network structure can fundamentally reshape eco-evolutionary dynamics and promote cooperation in complex systems.}

\section{\label{sec:level2}Mathematical model}
In the framework of ecological public goods \cite{wakano2009spatial,hauert2006evolutionary,hauert2008ecological}, we consider a well-mixed population consisting of cooperators and defectors with densities $u$ and $v$. Interactions occur in randomly formed groups of size $N$. The multiplication factor of the public good is denoted by $r>0$, and interactions may fail with probability $z=1-u-v$, representing ecological free space. The effective group size is therefore $S=(u+v)N$.

For a defector in a group of size $S$ facing $m$ cooperators, the expected payoff is
\[
P_D(S)=\frac{r}{S}\sum_{m=0}^{S-1}m
\binom{S-1}{m}
\left(\frac{u}{u+v}\right)^m
\left(\frac{v}{u+v}\right)^{S-1-m},
\]
while the payoff of a cooperator is
\(
P_C(S)=P_D(S)+\frac{r}{S}-1 .
\)

Averaging over all possible group sizes yields the expected fitness of defectors and cooperators
\begin{equation}
\label{eq:fDfC}
f_D=\frac{ru}{1-z}\left(1-\frac{1-z^N}{N(1-z)}\right),
\qquad
f_C=f_D-F(z),
\end{equation}
where
\(
F(z)=1+(r-1)z^{N-1}-\frac{r}{N}\frac{1-z^N}{1-z}.
\)

Population densities evolve according to reproductive success. Since reproduction requires available ecological space, it occurs at a rate $z(f_{C,D}+b)$, {\color{black}where $b$ denotes a baseline birth rate that maintains population viability even when payoff contributions are small. Additionally, individuals die at a constant rate $d$.} Combining reproduction and mortality yields the eco-evolutionary dynamics in a single well-mixed population as
\begin{equation}
\label{single_system}
\begin{aligned}
\dot{u}&=u\left(z(f_C+b)-d\right),\\
\dot{v}&=v\left(z(f_D+b)-d\right).
\end{aligned}
\end{equation}

\medskip

The detailed construction of the model \eqref{single_system} is described in the SI.1. In order to incorporate spatial structure, we embed the system in a complex network of $M$ nodes, each representing a local eco-evolutionary population described by Eq.~\ref{single_system}. Individuals can move between connected nodes, forming a metapopulation framework \cite{RobertMay1972}. \textcolor{black}{Let $A_{ij}$ denote the adjacency matrix of the network and $k_i=\sum_j A_{ij}$ the degree of node $i$. We consider a simple and undirected network, namely among any couple of nodes there can be one or no links at all, moreover the links are reciprocal, hence $A_{ij}=1$ if and only if there is an undirected link connecting node $i$ and node $j$, otherwise we set $A_{ij}=0$. Diffusion between nodes follows Fick's law and depends on density differences across network edges. For a generic variable $x_i$, the diffusive flux can be written as
\(
\sum_j A_{ij}(x_j-x_i),
\)
which is equivalent to Laplacian coupling
\(
\sum_j L_{ij}x_j,
\)
where $L_{ij}=A_{ij}-\delta_{ij}k_i$ is the network Laplacian. A system that evolves solely according to this diffusion rule (i.e., $\dot{x}_i =\sum_j L_{ij}x_j\,$) tries to locally equalize densities in connected nodes \cite{barrat2008dynamical}.}

Including this diffusion term yields the networked eco-evolutionary dynamics
\begin{equation}
\label{network_system}
\begin{aligned}
\dot{u_i}&=u_i\left(z_i(f_C+b)-d\right)+\epsilon\sum_{j=1}^{M}L_{ij}u_j,\\
\dot{v_i}&=v_i\left(z_i(f_D+b)-d\right)+\kappa\epsilon\sum_{j=1}^{M}L_{ij}v_j.
\end{aligned}
\end{equation}

Here $u_i$ and $v_i$ denote the densities of cooperators and defectors at node $i$, with $z_i=1-u_i-v_i$. The parameters $\epsilon$ and $\kappa\epsilon$ represent the diffusion rates of cooperators and defectors, respectively, while $\kappa$ denotes the ratio between the diffusion mobility of defectors and cooperators. This formulation captures the interplay between eco-evolutionary dynamics and diffusion on complex networks.

In the subsequent sections, we systematically analyze the emergence of Turing instability~\cite{Turing1952,nakao2010turing} in the network-structured eco-evolutionary system described by~\eqref{network_system}. {\color{black}The latter is a general mechanism capable of explaining the emergence of heterogeneous solutions from an initially homogeneous state through diffusion-driven destabilization. In the present work, we investigate this mechanism by performing a linear stability analysis around the stable fixed point of the isolated system ~\eqref{single_system}, and then explore the conditions under which diffusion-driven instabilities arise depending on the network topology and diffusion parameters used in ~\eqref{network_system}.} This instability disrupts the homogeneous coexistence of cooperators and defectors, leading to the spontaneous emergence of spatially heterogeneous patterns across the network nodes. This mechanism provides a deeper understanding of how spatial structure drives pattern formation and evolutionary transitions in eco-evolutionary dynamics.

\begin{figure}[htp]
		\centering
        \includegraphics[width=\columnwidth]{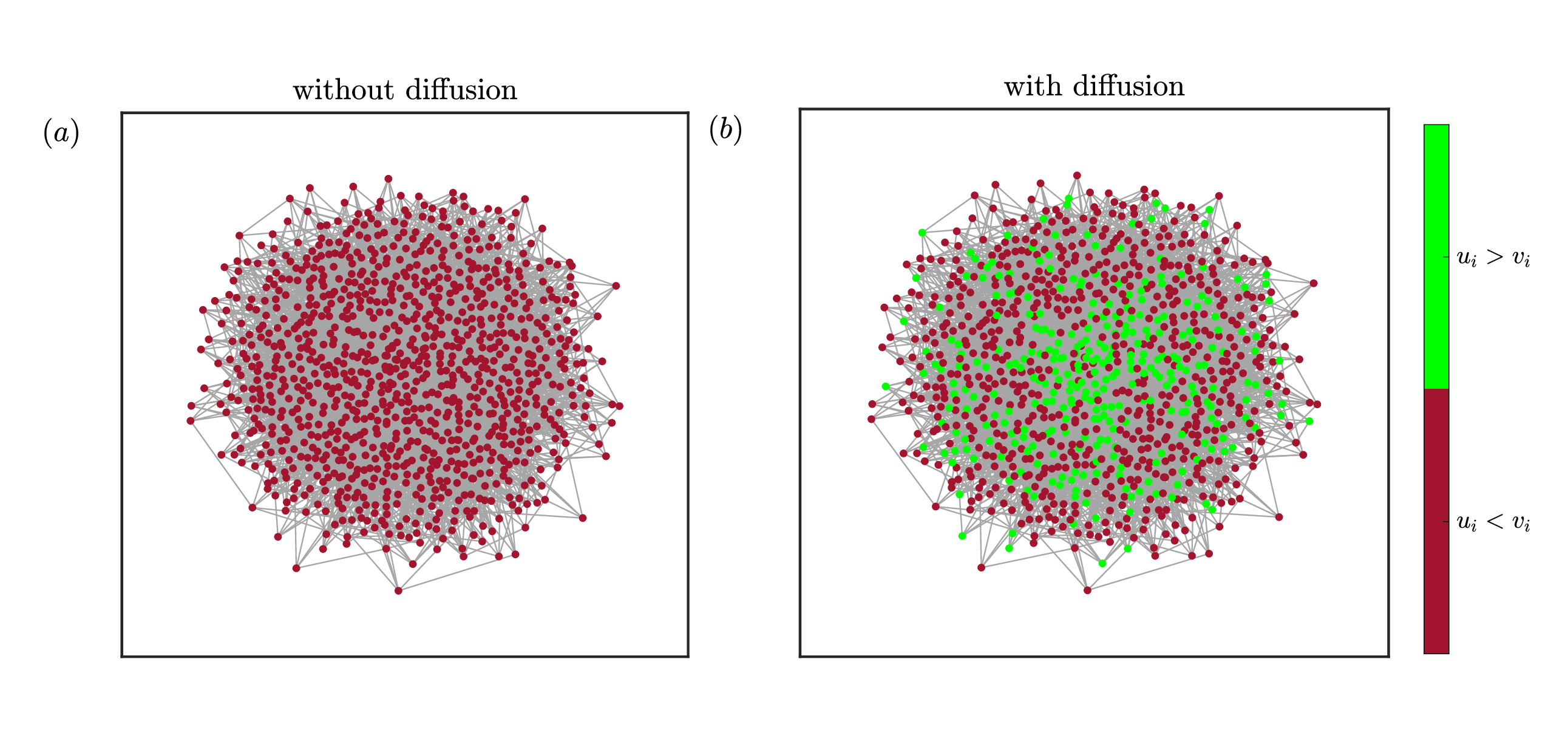} 
		\caption{
\textbf{Impact of asymmetric agents' mobility:} We report the result of a simulation of the ecological public goods game where agents interact and diffuse in a scale-free network (BA) of $1000$ nodes and average degree $<k> = 10$. $(a)$ shows the emerging network structure when the players cannot diffuse, i.e., $\epsilon=0$ in~\eqref{network_system}. One can observe that every node becomes a defector-dominated zone, {\color{black}$u_i<v_i$} (deep brown nodes). On the other hand, by inducing the diffusion ability, i.e., $\epsilon >0$ and $\kappa >0$, we can appreciate (see $b$) the existence of zones where cooperators can outweigh defectors, {\color{black}$u_i>v_i$} (green nodes); let us observe that those nodes have a relatively large degree, while defectors concentrate themselves, {\color{black} $v_i>u_i$}, in peripheral nodes, i.e., with smaller degree.}
		\label{network}
     \end{figure}

{\color{black} By anticipating the following analysis, we report in Fig.~\ref{network} the outcome of a representative simulation where cooperators and defectors play the ecological public goods game in each node of a scale-free network built using the Barabási-Albert algorithm (BA)~\cite{BarabasiAlbert2002}. In the absence of diffusion (Fig.~\ref{network}$(a)$), the dynamics in each node is ruled by~\eqref{single_system}, and the local densities converge to the coexistence equilibrium $u_i=u^*$ and $v_i=v^*$ for all nodes, with $u^*<v^*$. Thus, every node remains defector-prone (brown nodes, represented by $u_i<v_i$), reflecting the defector-dominated coexistence regime. Once diffusion is introduced (see Fig.~\ref{network}$(b)$), the homogeneous coexistence state becomes unstable and spatial heterogeneity emerges. In particular, a large number of high-degree nodes accumulate a larger fraction of cooperators (green), thereby reversing the local balance to $u_i>v_i$ in those nodes. This diffusion-induced stratification is further investigated in the subsequent section through numerical results and a mean-field analysis. Although defectors dominate globally, mobility asymmetry is associated with enhanced local cooperation in highly connected nodes, increasing local abundance and promoting the spread of cooperation across the network. A comprehensive overview of the temporal effect of agent mobility across the chosen network topology over time is provided in Supporting Information, Section~7, Movie~S1.}
\section{Results}
{\color{black}This section analyzes the main result shown in Fig.~\ref{network} and investigates the impact of the model parameters, $r$, $\kappa$, and $N$, as well as the network topology, on the system dynamics. Throughout this work, we focus on parameter regimes where the isolated system exhibits a defector-dominated coexistence stable and stationary equilibrium ($u^*<v^*$), allowing us to investigate whether asymmetric mobility and network heterogeneity can enhance cooperation beyond the baseline state.}

\subsection{Breaking the stable co-existence of competitive individuals}

{\color{black}{The eco-evolutionary public-goods framework without mobility, ~\eqref{single_system}, exhibits a stationary coexistence of cooperators and defectors at $(u^{*},v^{*})=(0.020078,0.102254)$ for the parameter set $r=4$, $N=30$, $d=1.3$, and $b=1$ (see SI.2, Figs.~S1-S2 for the stability analysis of the aspatial system). In this regime, defectors dominate, consistent with the classical outcome of public-goods games, with $ v^*$ nearly five times larger than $ u^*$ (horizontal reference lines in Fig.~\ref{scalefree1}($a$)). {\color{black}{The parameter values considered here are chosen intentionally so that the baseline system remains defector-dominated while still allowing Turing instability and diffusion-driven pattern formation.}}

A qualitatively different behavior emerges once mobility is introduced, with cooperators and defectors diffusing at different rates. The system is embedded in a scale-free network generated using the Barabási–Albert algorithm with $M=1000$ nodes and average degree $\langle k\rangle=10$~\cite{BarabasiAlbert2002,barabasi2003scale}, where each node hosts the local eco-evolutionary dynamics described by \eqref{single_system}. For diffusion parameters $\epsilon=0.01$ and mobility ratio $\kappa=100$, small heterogeneous perturbations applied to the homogeneous equilibrium of \eqref{network_system} evolve toward a spatially heterogeneous configuration, breaking the uniform coexistence of cooperators and defectors.
 \begin{figure}[htp]
		\centering
        \includegraphics[width=\columnwidth]{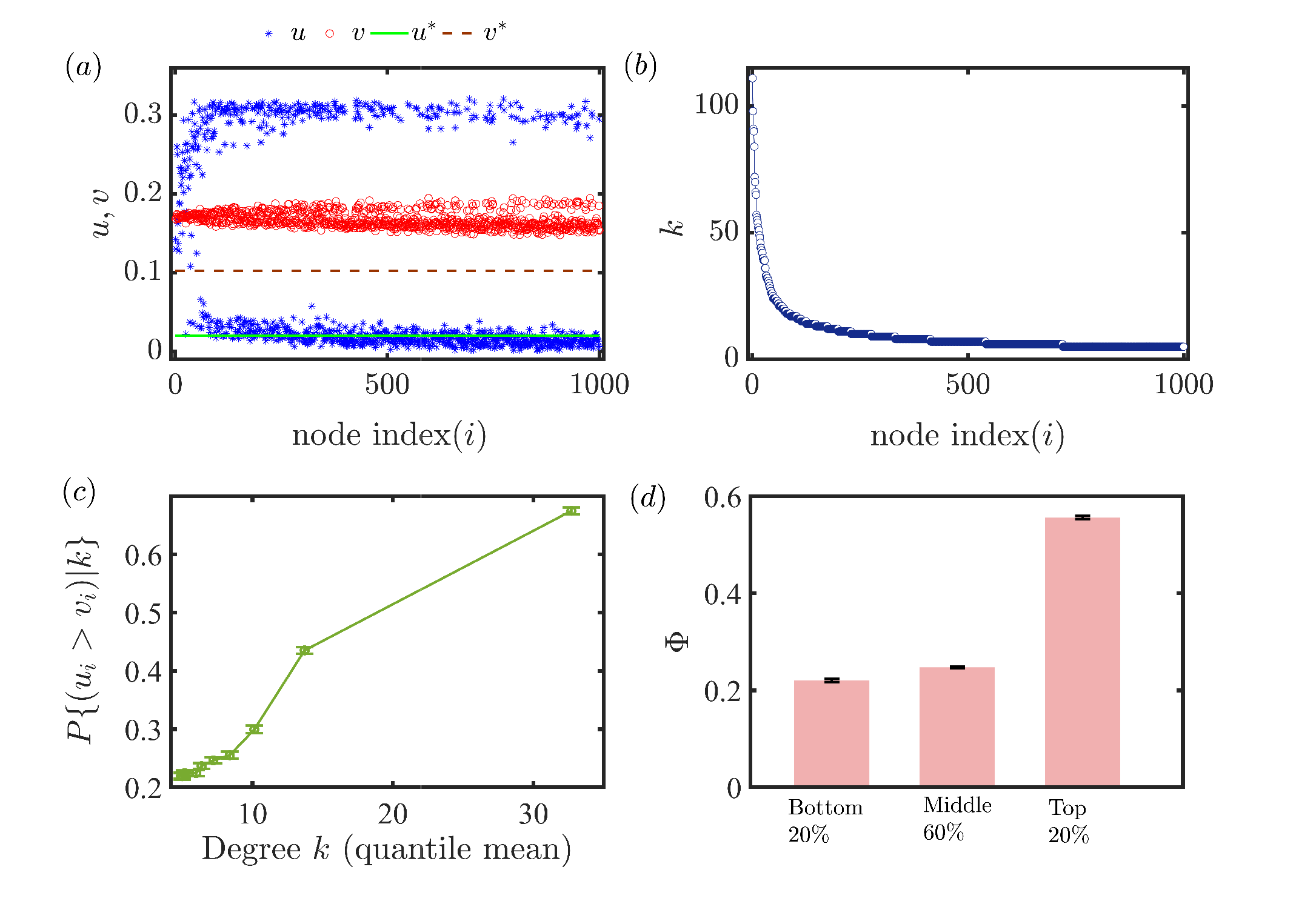} 
		\caption{\textcolor{black}{\textbf{Diffusion-induced heterogeneous cooperation patterns in a scale-free network.} (a) Stationary densities of cooperators $u_i$ (blue) and defectors $v_i$ (red) across nodes ordered by decreasing degree. Horizontal lines denote the homogeneous equilibrium values $(u^*,v^*)$ of the aspatial system. Diffusion with asymmetric mobility $(\epsilon=0.01, \kappa=100)$ produces strong spatial heterogeneity, with several nodes exhibiting cooperative dominance ($u_i>v_i$). (b) Degree distribution of the Barabási–Albert network with $M=1000$ nodes and average degree $\langle k\rangle=10$, showing the characteristic heavy-tailed connectivity of the scale-free topology. (c) $P\{(u_i>v_i)|k\}$, Probability of cooperator-dominated nodes $(u_i>v_i)$ as a function of node degree. Nodes are grouped using quantile-based binning, and the horizontal axis represents the mean degree within each bin. Points denote averages over $100$ realizations of network structure and initial conditions; error bars indicate the standard error. (d) Degree-stratified fraction $(\Phi)$ of cooperator-dominated nodes in the bottom $20\%$, middle $60\%$, and top $20\%$ of the degree distribution. Values represent averages over $100$ realizations, with standard error bars. Highly connected nodes exhibit a substantially larger probability of cooperative dominance than peripheral nodes.}}
		\label{scalefree1}
     \end{figure}
The temporal development of this instability is illustrated in SI Movie~1. Initially, node densities remain close to $(u^*,v^*)$, and spatial differentiation is minimal. As time progresses, diffusion amplifies small perturbations about the homogeneous state. Nodes with higher connectivity gradually exhibit increasing cooperative densities, deviating from $u^*$. The system eventually relaxes to a stationary heterogeneous configuration around $t=1000$, shown in Fig.~\ref{scalefree1}($a$), where nodes are ordered by decreasing degree. The diffusion-driven instability produces pronounced heterogeneity: approximately $35\%$ of nodes become cooperator-dominated ($u_i>v_i$), despite the isolated systems favoring defectors. The average asymptotic densities increase to $\langle v\rangle=0.167271$ and $\langle u\rangle=0.102184$, compared with the aspatial equilibrium values $v^*=0.102254$ and $u^*=0.020078$. Notably, the increase in cooperators is substantially larger than that of defectors, $(\langle u\rangle-u^*)\gg(\langle v\rangle-v^*)$, indicating that asymmetric mobility strongly amplifies cooperative abundance.

The emergence of these heterogeneous patterns is closely related to the network structural heterogeneity. The degree distribution in Fig.~\ref{scalefree1}($b$) reflects the heavy-tailed connectivity of the scale-free topology. Visual inspection of Fig.~\ref{scalefree1}($a$) suggests that nodes with larger degrees are more likely to exhibit cooperative dominance. This relationship is quantified by measuring the fraction of nodes satisfying $(u_i>v_i)$ as a function of degree (Fig.~\ref{scalefree1}($c,d$)). In fig.~\ref{scalefree1}($c$), nodes are grouped using quantile-based binning of the degree distribution, and the horizontal axis represents the mean degree within each bin. The plotted values, therefore, correspond to the probability that a node is cooperator-dominated within each degree class. Error bars denote the standard error computed across multiple realizations of both network structure and initial conditions. Because the degree axis represents quantile averages rather than individual node degrees, the largest value corresponds to the mean degree of the highest quantile group rather than the maximum degree present in the network.

Fig. \ref{scalefree1}$(d)$ provides a complementary comparison by grouping nodes into three percentile classes: the bottom $20\%$, the middle $60\%$, and the top $20\%$ of the degree distribution. The fractions $(\Phi)$ of cooperative dominance are again averaged over multiple realizations with standard error bars. Both panels show that the probability of cooperative dominance increases with node degree; highly connected nodes (in the top $20\%$) exhibit more than twice the fraction of cooperative dominance observed among peripheral nodes.

The dependence of cooperation on connectivity is further quantified by evaluating the Pearson correlation between node degree and the binary cooperation indicator $C_i=\mathbf{1}(u_i>v_i)$. Averaged over multiple realizations, a positive correlation ($\rho\approx0.17$, $p<10^{-5}$) is obtained, confirming that nodes with higher connectivity are statistically more likely to sustain cooperative dominance. Logistic regression yields the same conclusion: the probability of cooperative dominance increases significantly with node degree. Importantly, this relationship is probabilistic rather than deterministic: while some hubs remain defector-dominated, the likelihood of cooperative dominance increases systematically with node connectivity.

These results indicate that network connectivity plays a key role in shaping the eco-evolutionary dynamics. Nodes with higher degrees experience stronger diffusive coupling with the rest of the network and interact with more neighboring populations. Under asymmetric mobility, this enhanced coupling promotes the local accumulation of slowly diffusing cooperators by increasing the probability that highly connected nodes become cooperator-dominated. {\color{black}This mechanism is consistent with the degree-based mean-field analysis presented later, which shows that increasing node degree increases the effective coupling strength in the reduced dynamics and thereby shifts the local stationary state toward larger cooperator density.}

The robustness of the Turing mechanism is tested by repeating the numerical experiments on networks with different topologies, including small-world (Watts–Strogatz) and random (Erd\H{o}s-R\'enyi) graphs. The corresponding results (SI.3, Figs.~S3 and~S4, and SI Movies 2 and 3) show that diffusion destabilizes homogeneous coexistence and produces heterogeneous spatiotemporal patterns across these structures. Although these networks do not exhibit the strong degree of heterogeneity characteristic of scale-free networks, finite realizations still display moderate variability in node connectivity. Consistent with the behavior observed in the scale-free case, nodes with comparatively larger degrees in these networks are also more likely to exhibit cooperative dominance, indicating that the degree-dependent mechanism promoting cooperation is not restricted to scale-free topologies.

Overall, these results demonstrate that asymmetric mobility and heterogeneous network connectivity jointly reshape the eco-evolutionary dynamics of the public-goods game. Diffusion destabilizes the homogeneous equilibrium and promotes the emergence of cooperative clusters, which are preferentially associated with highly connected nodes, providing a structural pathway for cooperation to persist in competitive environments. {\color{black} We note that the parameter regimes considered here yield stationary heterogeneous patterns. Different parameter choices may also produce oscillatory or chaotic diffusion-driven states, as reported in earlier work \cite{wakano2009spatial}, and remain interesting directions for future investigation.}}}

           \begin{figure*}[htp!]
		\centering
        \includegraphics[width=0.7\linewidth]{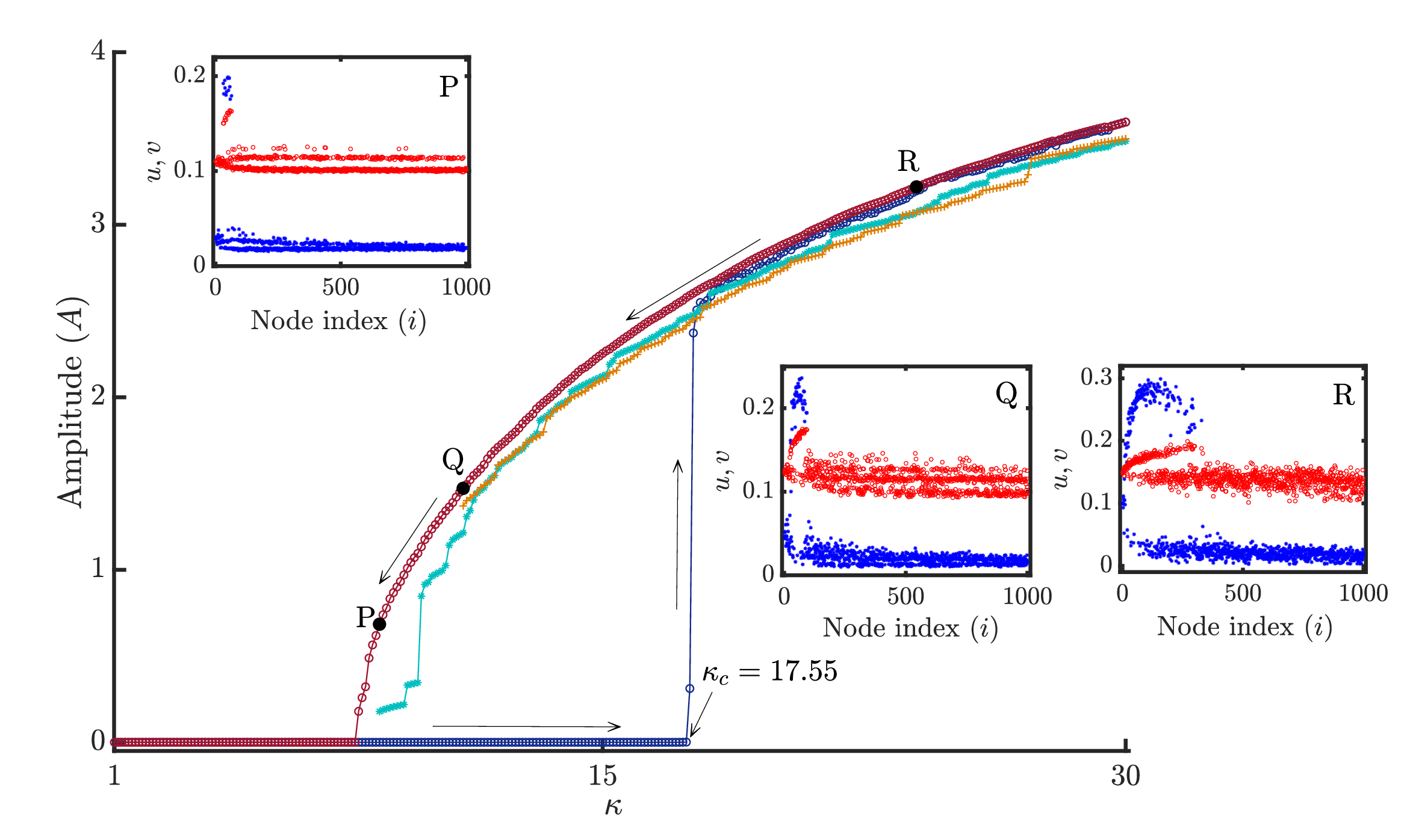} 
		\caption{
\textbf{Amplitude bifurcation reveals hysteresis and multistability in mobility-induced pattern formation.}
This figure shows the results of an amplitude bifurcation analysis performed on the networked eco-evolutionary public goods system \eqref{network_system} implemented on a scale-free (BA) network with $M=1000$ nodes and mean degree $\langle k \rangle = 10$. The system evolved to a stationary equilibrium already at $t = 1000$ time steps for each value of $\kappa$, representing the relative mobility of defectors ($v$) to cooperators ($u$), while other parameters were held constant at values that maintained stable coexistence in the isolated system. The main panels show the network-averaged amplitude $A$, which measures the deviation from spatial homogeneity, as a function of $\kappa$. Blue hollow circles represent adiabatic forward continuation (increasing $\kappa$), where at each step the system is allowed to evolve from a stable fixed point with small perturbations, capturing the emergence of instability. Red hollow circles denote adiabatic backward continuation (decreasing $\kappa$), starting from a patterned state and tracking the persistence of spatial structure. A critical threshold $\kappa_c \approx 17.55$ marks the onset of pattern formation through a sharp increase in $A$. The system exhibits hysteresis: decreasing $\kappa$ fails to immediately suppress the pattern, indicating bistability and the presence of multiple attractors—a hallmark of subcritical bifurcation. The figure also includes thin cyan and orange-starred curves that illustrate auxiliary forward and backward continuations under small perturbations, demonstrating sensitivity to initial conditions and network realizations. Insets P, Q, and R show the stationary node-wise abundances of cooperators ($u$, blue) and defectors ($v$, red) at $\kappa = 8.6$, $11$, and $24$, respectively. Even at subcritical values below $\kappa_c$, strong spatial heterogeneity persists, illustrating the importance of network structure and history dependence.
}
		\label{amplitude}
     \end{figure*}

\subsection{Hysteresis and multistability in mobility-induced pattern formation}

To understand how diffusion asymmetry between cooperators and defectors drives spatial pattern formation in the proposed eco-evolutionary public goods system, we conduct an amplitude analysis by gradually increasing and decreasing the diffusion ratio $\kappa$, keeping all the other parameters of individual nodes and network structure the same as in Fig.~\ref{scalefree1}. For each value of $\kappa$, we compute the network-wide amplitude $A$, which measures how far the current state deviates from spatial homogeneity:
$$
A = \left[ \sum_{i=1}^{M} \left( (u_i - u^*)^2 + (v_i - v^*)^2 \right) \right]^{1/2},
$$
where $u^*$ and $v^*$ are the steady state densities of the cooperators and defectors, respectively, without diffusion. Let us note that we did not normalize the amplitude with respect to the network size, as we deal with fixed sizes in this work. The amplitude $A$ acts as an indicator of spatial pattern formation: values close to zero indicate uniform co-existence, while higher values reflect increasing heterogeneity, i.e., the emergence of localized cooperation or defection subgroups.

Figure~\ref{amplitude} displays the results of the amplitude analysis. During forward continuation (blue circles), $\kappa$ is increased gradually, and the system is allowed to relax to stationarity after each increment. The amplitude remains close to zero until a sharp transition near $\kappa_c \approx 17.55$ emerges, beyond which $A$ rises abruptly, signaling the onset of diffusion-driven spatial pattern formation. Further increases in $\kappa$ strengthen the heterogeneity (see inset R).

A different behavior appears during backward continuation (red circles), where $\kappa$ is decreased after reaching the patterned regime. The system does not immediately return to the homogeneous state but instead remains trapped in patterned configurations even for $\kappa$ well below the forward threshold (insets P and Q). This mismatch between forward and backward trajectories forms a hysteresis loop, indicating a subcritical bifurcation and multistability between homogeneous and patterned states.

Additional branches (cyan and orange starred curves) are obtained by initializing the system with patterned states 
$P$ and $Q$, and performing forward and backward continuations. In these cases, the amplitude increases before the critical value of the main branch, highlighting the sensitivity of pattern emergence to initial conditions and network realization. Similar multibranch structures have been reported in networked Turing systems such as the Mimura–Murray model \cite{nakao2010turing}.

The insets of Fig.~\ref{amplitude} display representative stationary patterns. Even for $\kappa=8.6$ and $\kappa=11$ (corresponding to points P and Q), which lie below the forward critical threshold, strong spatial fluctuations persist, confirming the bistable coexistence of uniform and patterned solutions. Panel R, corresponding to $\kappa = 24$, reveals a well-developed spatial pattern: groups of high cooperator density emerge in specific zones of the network, particularly among high-degree nodes that are less vulnerable to invasion by fast-moving defectors.

Thus, the bifurcation structure reveals how mobility asymmetry can promote cooperation through spatial differentiation. Rapidly diffusing defectors homogenize the network and deplete shared resources, while slowly diffusing cooperators remain localized, forming protected cooperative clusters. Such mechanisms resemble observations in microbial systems where restricted mobility—e.g., in biofilms or viscous environments—allows cooperative strains producing public goods to persist locally despite global exploitation \cite{cremer2019chemotaxis}. Similar spatial constraints also promote cooperation in many social and ecological systems.

           \begin{figure}[htp]
		\centering
        \includegraphics[width = \columnwidth]{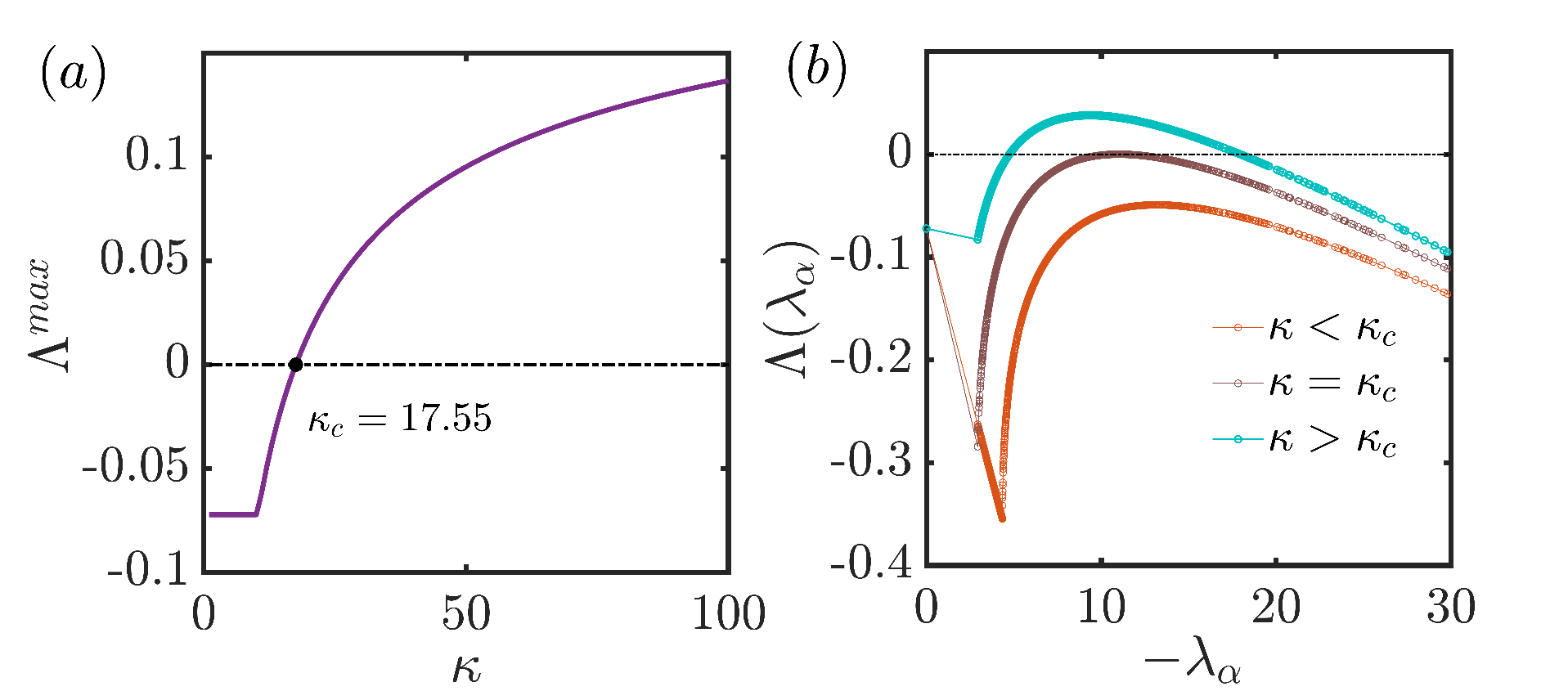} 
		\caption{
\textbf{Analytical characterization of the critical diffusion ratio $\kappa_c$}: 
($a$) Maximum growth rate $\Lambda^{\mathrm{max}}$ as a function of the mobility ratio $\kappa$. The curve crosses zero at $\kappa_c \approx 17.55$, which marks the threshold of diffusion-induced instability (marked solid black circle). For $\kappa < \kappa_c$, all perturbations decay and the homogeneous coexistence state remains stable, while beyond this point, instabilities emerge. 
($b$) Dispersion relation $\Lambda(\lambda_\alpha)$ for three representative values of $\kappa$: below the threshold ($\kappa < \kappa_c$, orange), exactly at the threshold ($\kappa = \kappa_c$, brown), and above the threshold ($\kappa > \kappa_c$, cyan). Below $\kappa_c$, all transverse modes are stable; at $\kappa = \kappa_c$ the leading mode becomes marginally stable; and for $\kappa > \kappa_c$ several modes cross into the positive regime, indicating exponential amplification of spatial perturbations. Together, these panels confirm analytically the onset of Turing instability driven by the mobility of strategy-driven individuals.
}
		\label{analytics}
     \end{figure}

\subsection{The critical diffusion ratio $\kappa_c$} \label{3c}

\textcolor{black}{The onset of diffusion–driven instability in the eco–evolutionary
networked system~\eqref{network_system} is determined by the critical diffusion ratio $\kappa_c$. The analytical derivation of this threshold, obtained through the linear stability analysis of the homogeneous coexistence equilibrium $(u^*,v^*)$ and the spectral decomposition of the network Laplacian, is presented in the Materials and Methods
section~\ref{kappa_c}. Here, we summarize the resulting instability
condition and its dynamical consequences.}

The analysis yields an explicit expression for the critical diffusion ratio,
\begin{equation}
\label{analytical_kappa1}
\kappa_c =
\frac{f_u g_v - 2 f_v g_u + 2\sqrt{f_v g_u \,(f_v g_u - f_u g_v)}}{f_u^2},
\end{equation}
where $f_u, f_v, g_u$, and $g_v$ denote the elements of the Jacobian
matrix of the local eco–evolutionary dynamics evaluated at the
homogeneous coexistence equilibrium $(u^*,v^*)$ (see~\eqref{eq:Jalpha}
).

The analytical prediction is illustrated in Fig.~\ref{analytics}$(a)$,
which shows the maximum $\Lambda^{\mathrm{max}}$ of the
dispersion relation as a function of the mobility ratio $\kappa$.
For small values of $\kappa$, $\Lambda^{\mathrm{max}}$ remains negative,
indicating that all perturbations decay and the homogeneous coexistence
state $(u^*,v^*)$ is linearly stable. As $\kappa$ increases,
$\Lambda^{\mathrm{max}}$ progressively approaches the stability boundary and eventually crosses zero at $\kappa_c \approx 17.55$ (see the black dot in Fig.~\ref{analytics}$(a)$).
This crossing point marks the critical threshold beyond which the homogeneous equilibrium loses stability and diffusion–driven spatial heterogeneity can emerge.

Additional insight is provided by Fig.~\ref{analytics}$(b)$, which displays the dispersion relation $\Lambda(\lambda_\alpha)$ (see~\eqref{eq:reldispmax}) as a function of $-\lambda_\alpha$, the eigenvalues of the Laplace matrix, for three representative values of the diffusion ratio. When $\kappa < \kappa_c$ (the sea blue curve in Fig.~\ref{analytics}$(b)$), the entire curve lies below zero, implying that all perturbation modes associated with the Laplacian eigenvalues decay and the homogeneous
solution remains stable. At $\kappa = \kappa_c$ (see brown curve in Fig.~\ref{analytics}$(b)$), the leading mode
horizontally touches the zero line, indicating marginal stability of the system. For $\kappa > \kappa_c$ (orange curve in Fig.~\ref{analytics}$(b)$), part of the dispersion relation becomes positive, meaning that a subset of Laplacian eigenmodes induce exponential growth of any small perturbation and destabilizes the uniform coexistence state.

Taken together, Fig.~\ref{analytics}$(a)$ and Fig.~\ref{analytics}$(b)$ confirm that the transition at $\kappa_c$ corresponds to a genuine diffusion–driven instability. Once the mobility asymmetry between cooperators and defectors exceeds this threshold, the homogeneous eco–evolutionary equilibrium becomes unstable, and spatially heterogeneous patterns can develop across the network.

{\color{black} \subsection{Mean-field approximation} \label{2.4}
To explain the numerical patterns observed at large mobility asymmetry and to unravel the degree-dependent organization reported in Fig.~\ref{scalefree1}, we employ a degree-based mean-field approximation similar to those used for reaction–diffusion dynamics on complex networks \cite{nakao2010turing}. In this framework, the detailed interactions between a node and its neighbors are replaced by coupling to global mean fields representing the collective state of the network \cite{nakao2009diffusion,colizza2007reaction,colizza2008epidemic,barrat2008dynamical}. The derivation of this method and the computation of the mean fields are provided in the Materials and Methods section~\ref{meanfield_derive}.

Under this approximation, the network dynamics reduces to the degree-dependent system
\begin{equation}\label{meanfield_eq_main_text}
\begin{aligned}
\dot{u} &= u[(1-u-v)(f_C+b)-d]+\beta(\langle u\rangle-u), \\
\dot{v} &= v[(1-u-v)(f_D+b)-d]+\kappa\beta(\langle v\rangle-v),
\end{aligned}
\end{equation}
where $\langle u\rangle$ and $\langle v\rangle$ denote the global mean densities of cooperators and defectors, and the effective coupling parameter $\beta=\varepsilon k$ is proportional to the node degree $k$.

 \begin{figure}[htp]
		\centering
        \includegraphics[width=\columnwidth]{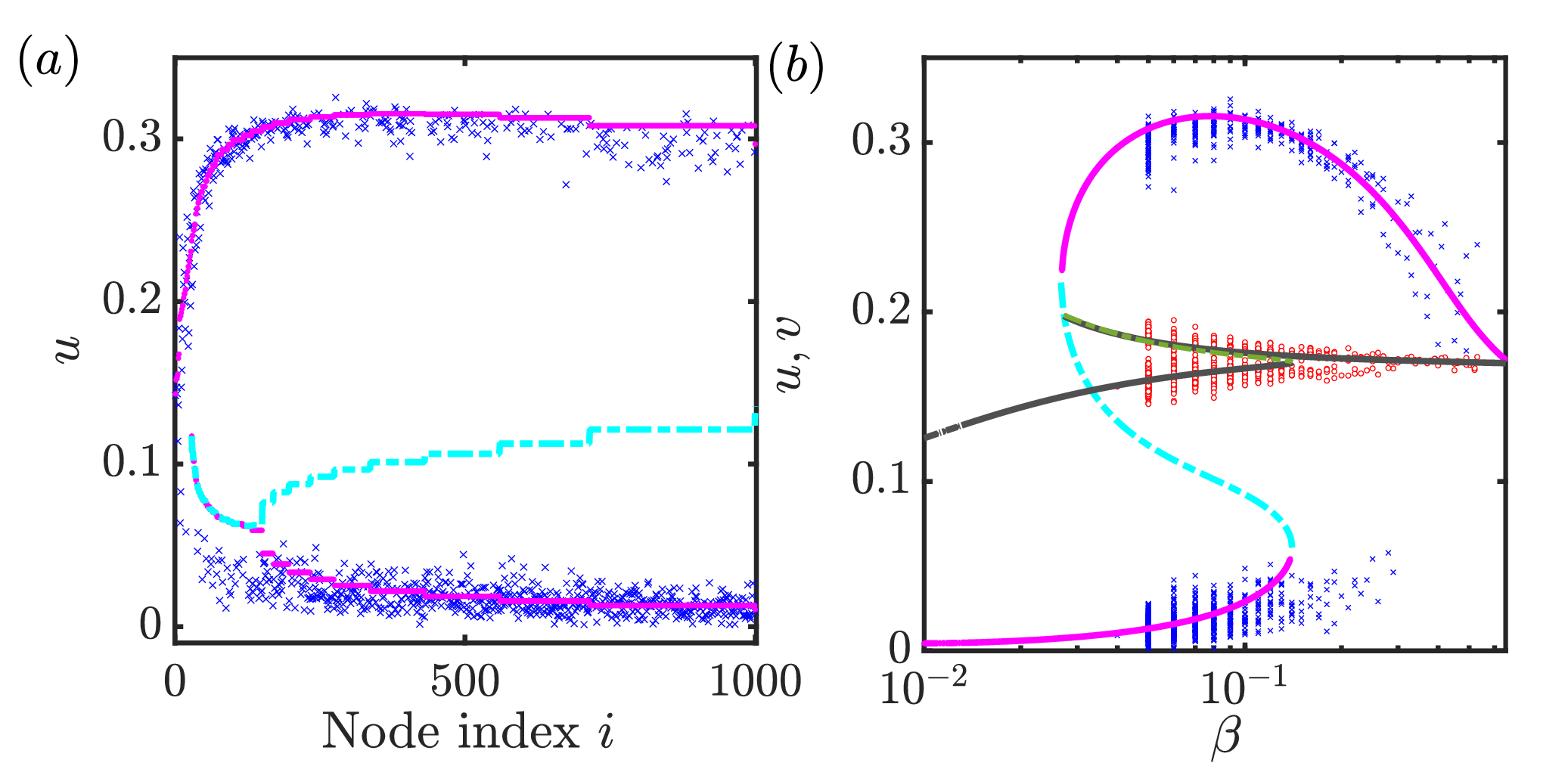} 
		\caption{ \color{black}
        \textbf{Degree-Based Mean-Field Approximation and Bifurcation Structure:}$(a)$ shows the degree-resolved stationary cooperator densities $u_i$ plotted against the ordered node index (ordered by decreasing degree) for $\kappa = 100$ and $\epsilon=0.01$. Blue crosses denote the cooperator densities from full network simulations, while the magenta (cyan) curves represent the corresponding stationary stable (unstable) solutions obtained from the reduced mean-field equation. $(b)$ shows the bifurcation diagram of the reduced mean-field system as a function of the effective coupling parameter $\beta = \varepsilon k$ for $\epsilon=0.01$. Magenta (cyan) curves denote stable (unstable) stationary branches of cooperators, while dark (light) green curves represent stable (unstable) branches of defectors. Blue crosses and red circles depict the cooperator and defector densities from network simulation for $\kappa=100$ and $\epsilon=0.01$. A saddle-node bifurcation generates two coexisting stable states over an intermediate range of $\beta$, explaining the emergence of cooperative dominance over high-degree nodes and the multistable behavior observed in the network simulations.} 
		\label{meanfield_cal}
     \end{figure}

The bifurcation structure of this reduced system is shown in Fig.~\ref{meanfield_cal}($b$). Magenta (cyan) curves denote stable (unstable) stationary branches of the cooperator density, while dark (light) green curves represent stable (unstable) branches of the defector density. As the effective coupling parameter $\beta$ increases, the system undergoes a saddle-node bifurcation, giving rise to two coexisting stable states over an intermediate range of $\beta$. For small values of $\beta$, corresponding to weak diffusive coupling, the system remains close to the lower branch associated with the defector-dominated coexistence equilibrium. As $\beta$ increases beyond the saddle-node threshold, a second stable branch emerges that is characterized by substantially higher cooperator densities.

Because the effective coupling parameter scales linearly with node degree, this bifurcation structure translates directly into degree-dependent dynamics on the network. Fig.~\ref{meanfield_cal}($a$) compares the stationary cooperator densities obtained from full network simulations (red crosses) with the stationary branches predicted by the mean-field equations (magenta curves) evaluated at $\beta_{i}=\epsilon k_{i}$ for $\epsilon=0.01$ and $\kappa=100$. In the mean-field description, each node $i$ is therefore characterized by its own bifurcation parameter value determined by its degree. The stationary network pattern can thus be interpreted as a set of points distributed along the bifurcation diagram of the reduced system.

The comparison reveals a clear separation between low- and high-degree nodes. Nodes with small degree cluster near the lower stable branch corresponding to defector-dominated coexistence, whereas highly connected nodes populate the upper stable branch associated with enhanced cooperator density. Although some scattering of numerical data appears near the bifurcation points, the overall agreement between simulation results and mean-field predictions demonstrates that the heterogeneous network pattern is well explained by the bifurcation structure of a single node coupled to the global mean fields, with the effective coupling strength determined by the node degree.

Remarkably, the bifurcation structure also shows multistability. Over an intermediate range of the coupling parameter $\beta$, both stationary branches coexist as stable solutions. Consequently, nodes with similar degrees may converge to different stationary states depending on perturbations or initial conditions. This explains why, although the probability of cooperative dominance increases with node degree (Fig.~\ref{scalefree1}), not all hubs become cooperator-dominated in the network simulations. }

\section{Impact of the multiplication factor $r$ and diffusion ratio $\kappa$ on diffusion-driven instability}

To further understand how cooperative amplification and mobility asymmetry jointly shape the emergence of spatial patterns, we analyze the stability of the homogeneous state by calculating the maximum of the dispersion relation $\Lambda^{\mathrm{max}}$ in the two–dimensional parameter space defined by the public goods multiplication factor $r$ and the diffusion ratio $\kappa$.
The sign of $\Lambda^{\mathrm{max}}$ determines the stability of the homogeneous equilibrium of system~\eqref{network_system}: negative values correspond to stable homogeneous dynamics, whereas $\Lambda^{\mathrm{max}}>0$ signals diffusion-driven instability.

\begin{figure*}[htp]
		\centering
        \includegraphics[width=0.8\textwidth]{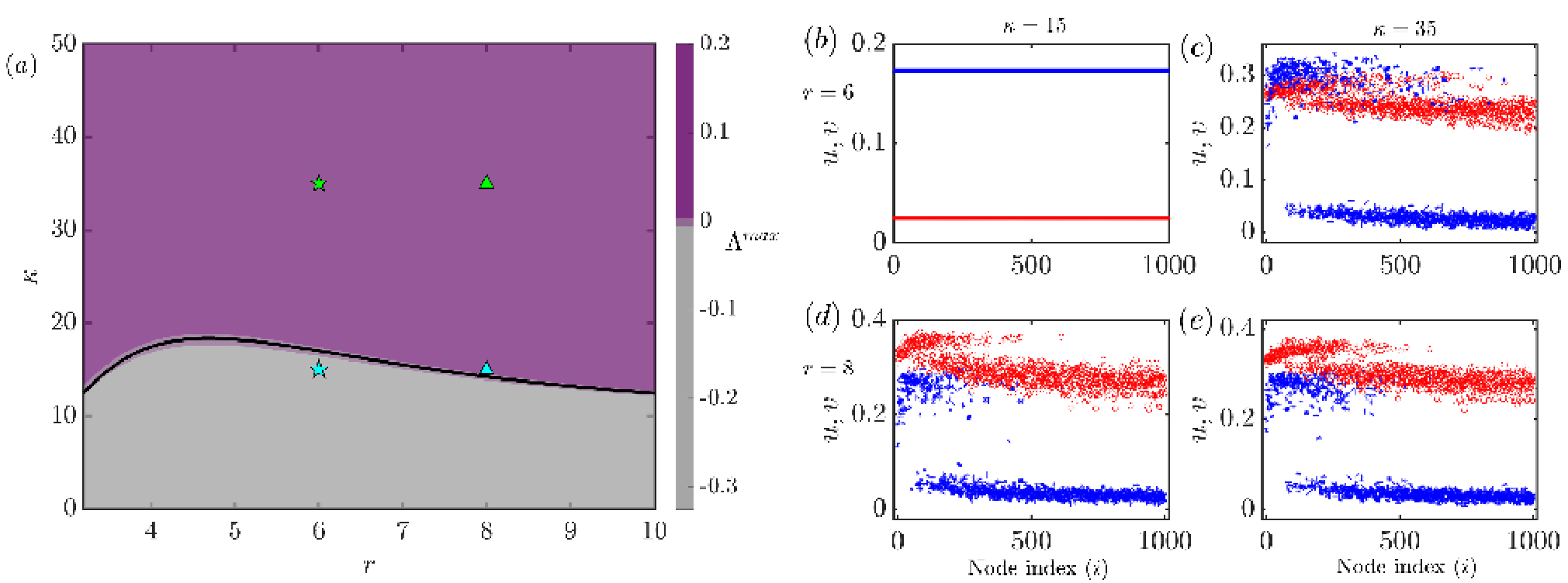} 
		\caption{
\textbf{Diffusion-induced instability in $(r,\kappa)$ parameter space.}
We show the maximum of the dispersion relation $\Lambda^{\mathrm{max}}$ as a function of $r$ and $\kappa$, as computed from the theory presented in the section~\ref{3c}. The latter indicates regions of stability and instability in the eco-evolutionary public goods system. In (a) purple regions correspond to unstable regimes where $\Lambda^{\mathrm{max}}(\kappa, r) > 0$ and spatial patterns emerge; grey regions indicate stable coexistence ($\Lambda^{\mathrm{max}}(\kappa,r) < 0$). The analysis is performed for $\epsilon = 0.01$, $N = 30$, $b = 1$, $d = 1.3$ on a scale-free network ($N=1000$, $\langle k \rangle=10$). Patterns are most likely to emerge in intermediate ranges of the public goods multiplication factor $r$ and diffusion ratio $\kappa$. The black curve represents the critical curve of $\kappa_c$, obtained from \eqref{analytical_kappa1}. The numerical validations for various combinations of $(r,\kappa)$, distinguished by different colors and markers, are shown in ($b$)-($e$). Here, cyan, and light green markers represent $\kappa=15,\text{and}\; 35$, respectively, whereas stars, and triangles correspond to $r=6,\text{and}\; 8$, respectively. 
Each panel shows $u$ (blue) and $v$ (red) for all $1000$ nodes. At low $\kappa$ (($b$)), the system remains homogeneous; as $\kappa$ increases (see ($c$)), spatial heterogeneity emerges with pronounced local fluctuations. Notably, higher $r$ values correlate with reduced cooperation levels ($(d)-(e)$), especially at high $\kappa$, highlighting how group-size dilution undermines cooperation in real-world public goods dilemmas.}
		\label{r_kappa_analytics}
     \end{figure*}

The analytical stability diagram is shown in Fig.~\ref{r_kappa_analytics}$(a)$. The black solid curve represents the critical diffusion ratio $\kappa_c(r)$ obtained from \eqref{analytical_kappa1}. This curve separates the stable regime, $\Lambda^{\mathrm{max}}<0$, from the instability region, $\Lambda^{\mathrm{max}}>0$, where diffusion destabilizes the homogeneous coexistence equilibrium and spatial heterogeneity may develop. The dependence of $\kappa_c$ on $r$ is non–monotonic: the threshold initially increases with $r$, reaches a maximum near $r\approx5$, and then decreases. Consequently, sufficiently large diffusion asymmetry destabilizes the homogeneous state over a broad range of multiplication factors.

To illustrate this analytical prediction, four representative parameter combinations were selected and marked in Fig.~\ref{r_kappa_analytics}$(a)$. Two values of the multiplication factor were considered, $r=6$ and $r=8$, together with two diffusion ratios, $\kappa=15$ and $\kappa=35$. The points $(6,15)$ and $(8,15)$ are indicated by cyan symbols (star and triangle), while $(6,35)$ and $(8,35)$ are represented by light–green symbols (star and triangle). The corresponding numerical simulations are shown in Fig.~\ref{r_kappa_analytics}$(b)$–$(e)$, where the stationary densities of cooperators ($u_i$) and defectors ($v_i$) are plotted across the $M=1000$ nodes of the network. For the parameter set $(6,15)$, located below the instability boundary, the system converges to a homogeneous state, as shown in Fig.~\ref{r_kappa_analytics}$(b)$. In contrast, the remaining three parameter combinations lie above the analytical threshold $\kappa_c(r)$ and produce spatially heterogeneous states, visible in Fig.~\ref{r_kappa_analytics}$(c)$–$(e)$.

These results confirm that the instability condition $\kappa>\kappa_c(r)$ predicted by the analytical dispersion relation accurately determines the emergence of heterogeneous cooperative patterns. The agreement between the analytical stability boundary and the numerical outcomes demonstrates that asymmetric mobility between cooperators and defectors destabilizes the homogeneous eco–evolutionary equilibrium and generates spatial organization across the network.

The results also reveal that increasing the multiplication factor $r$ does not always promote cooperation once spatial dynamics are present (Figs.~\ref{r_kappa_analytics} (d), (e)). In some regimes, larger $r$ reduces cooperative density because amplified group benefits increase defectors' exploitation. Thus, cooperation persists not simply through higher rewards but through spatial segregation enabled by mobility asymmetry. Taken together, the analytical stability map and numerical simulations show that cooperation emerges within a structured region of the $(r,\kappa)$ parameter space, where payoff amplification and differential mobility destabilize the homogeneous state and allow cooperative clusters to persist.
 
\section{Conclusions and discussions}
In this study, we introduced and analyzed an eco-evolutionary public goods game model on complex networks, focusing on the role of asymmetric diffusion between cooperators and defectors. By combining localized group interactions, ecological constraints, and network-structured mobility, the model provides a comprehensive framework to investigate how cooperation can emerge and persist in spatially structured populations.

\textcolor{black}{A central outcome of our investigation is that the faster diffusion of defectors relative to cooperators destabilizes homogeneous coexistence in which defectors dominate, leading to a symmetry-breaking transition and the emergence of a self-organized, cooperation-rich population. These patterns were validated numerically through amplitude bifurcation analysis and analytically supported by the dispersion relation, which together illustrate how heterogeneous network topology underlies the formation of spatially localized cooperative zones: highly connected nodes exhibit a higher probability of cooperative dominance, while peripheral nodes remain more susceptible to defector invasion.}

\textcolor{black}{To explain this mechanism, we developed a degree-based mean-field description of the network dynamics. The analysis shows that the effective coupling strength experienced by each node scales with its degree. Consequently, nodes with different degrees effectively operate at different positions along the bifurcation diagram of the reduced system. The bifurcation analysis reveals a saddle-node structure that produces bistability between a defector-dominated state and a cooperative state. This theoretical prediction accurately reproduces the heterogeneous patterns observed in the full network simulations and explains why cooperation tends to emerge preferentially in highly connected nodes. At the same time, the presence of multistability implies that transitions are probabilistic rather than deterministic, so that not all hubs necessarily become cooperator-dominated.}

Importantly, the system also exhibits hysteresis. Once spatial cooperation patterns are established, reversing the conditions that initially led to their formation does not necessarily return the system to its original state. This path dependence highlights the strong influence of initial conditions and perturbations on shaping long-term dynamics, revealing a landscape of alternative stable states. The agreement between numerical simulations and analytical predictions further reinforces the generality of these behaviors. \textcolor{black}{It is important to note that the present study focuses on representative parameter regimes that allow diffusion-driven instability in the baseline defector-dominated system. Although the analytical treatment captures the onset of instability and the degree-dependent organization of the stationary patterns, it does not provide a complete analytical characterization of the full parameter space nor a derivation of the increase in global average densities. Extending the theory in these directions remains an important topic for future work.}

A comprehensive exploration of the parameter space, including the synergy factor ($r$), group size ($N$), and diffusion asymmetry ratio ($\kappa$), reveals that cooperation is maximized through a nuanced balance among these factors. While a higher $r$ tends to promote cooperation, it can also increase susceptibility to defection without spatial support. Larger group sizes amplify collective benefit but may dilute individual payoffs ({\color{black}see SI.4, Fig.~S5}), and diffusion asymmetry broadens the range of instability, enabling richer pattern dynamics. Together, these parameters define a complex regime in which cooperation emerges through spatial differentiation rather than mean-field expectations.

From an eco-evolutionary perspective, our findings demonstrate that the interplay between asymmetric mobility and heterogeneous connectivity provides a powerful mechanism for promoting cooperation in structured populations. Such mechanisms are relevant for a variety of ecological and social systems, including microbial communities, collective resource management, and social interaction networks, where mobility and structural heterogeneity influence cooperative behavior.

In conclusion, our study underscores the need to incorporate spatial heterogeneity, mobility asymmetry, and network topology into models of eco-evolutionary dynamics. By combining large-scale simulations with analytical techniques, we show how structured populations can defy classical expectations, allowing cooperation to thrive even under ecologically adverse conditions. These findings offer practical insights into designing strategies for fostering cooperation in real-world systems, ranging from microbial ecosystems to social networks, by leveraging control over mobility, interaction structure, and group composition. Looking ahead, the generality of our framework invites verification across additional network classes and empirical interaction topologies, thus representing a promising avenue for future research, with the potential to further deepen our understanding of how cooperation emerges and persists under diverse ecological and social conditions. Let us conclude by emphasizing another possible direction, rooted in a more realistic diffusion process; indeed~\eqref{network_system} models a linear diffusion process, while rational agents would decide where to go or which nodes to avoid, for instance, cooperators could be less prone to move toward nodes with a larger density of defectors. This behavior can be included by replacing the linear diffusion terms in~\eqref{network_system} with a non-linear one, e.g., $\dot{u}_i \sim \epsilon \sum_j L_{ij} h(u_j,v_j)$ where $h$ is some nonlinear function encoding the rational choice of the players; a similar term can be inserted for the evolution rate of $v_i$.

\section{Materials and Methods}

{\color{black} \subsection{Linear stability analysis and critical diffusion ratio} \label{kappa_c}
To determine the onset of diffusion–driven instability discussed in Sec.~\ref{3c}, we perform a linear stability analysis of the networked eco–evolutionary
system ~\eqref{network_system} about the homogeneous coexistence equilibrium
$x^*=(u^*,v^*)$ of the isolated dynamics \eqref{single_system}.

We introduce small perturbations at each node, $\delta x_i=(\delta u_i,\delta v_i)^\top$, defined as $\delta u_i=u_i-u^*, \; \delta v_i=v_i-v^*$.
Substituting these expressions into ~\eqref{network_system} and retaining only
first–order terms yields
\begin{equation*}
\frac{d}{dt}\delta x_i = J_F(x^*)\,\delta x_i + \sum_{j=1}^{M} L_{ij}D\,\delta x_j ,
\end{equation*}
where $J_F(x^*)$ is the Jacobian matrix of the local eco–evolutionary
dynamics evaluated at $(u^*,v^*)$, i.e., 
$J_F(x^*)=
\begin{pmatrix}
f_u & f_v\\
g_u & g_v
\end{pmatrix}$, and \(
D=
\begin{pmatrix}
\epsilon & 0\\
0 & \kappa\epsilon
\end{pmatrix}
\) is the diffusion matrix describing the asymmetric mobility
of cooperators and defectors.

Because the Laplacian matrix $L$ is diagonalizable, the perturbations can be expanded in its eigenbasis. Let $\lambda_\alpha$ denote the Laplacian eigenvalues, and $\delta y_\alpha$ the amplitude of the perturbation along the eigenvector corresponding to mode $\alpha$. Because the Laplacian matrix $L$ is diagonalizable, the perturbations can be expanded in its eigenbasis. Let $\lambda_\alpha$ denote the Laplacian eigenvalues, and $\delta y_\alpha$ the coefficient of the perturbation along the eigenvector $\phi^{(\alpha)}$ associated with $\lambda_\alpha$. The perturbation modes then decouple as

\begin{equation}
\label{eq:Jalpha}
\frac{d}{dt}\delta y_\alpha
=
\left[J_F(x^*)+\lambda_\alpha D\right]\delta y_\alpha
\equiv J_\alpha \delta y_\alpha .
\end{equation}

The stability of each mode is determined by the eigenvalues of
the matrix $J_\alpha$. We therefore define the dispersion relation 
\begin{equation}
\label{eq:reldispmax}
\Lambda(\lambda_\alpha) = \max \Re\left[\sigma(J_\alpha)\right]\, ,  
\end{equation}
where $\sigma(J_\alpha)$ denotes the spectrum of $J_\alpha$. The homogeneous equilibrium is stable when $\Lambda(\lambda_\alpha)<0$ for all transverse modes
$\alpha>1$. Conversely, diffusion–driven instability occurs if $\Lambda(\lambda_\alpha)>0$ for some mode. In other words, The dispersion relation, $\Lambda(\lambda_\alpha)$ solves
\begin{equation}
\label{eq:reldisp}
\begin{array}{l}
\Lambda^2-\Lambda \left[\mathrm{tr}(J_F(x^*))+\epsilon \lambda_\alpha (1+\kappa)\right]+\det(J_F(x^*)) \\
     ~~~~~~~~~~~~~~~~~~ +\epsilon \lambda_\alpha (g_v+\kappa f_u)+\epsilon^2 \lambda_\alpha^2 \kappa=0\, ,
     \end{array}
\end{equation}
where to lighten the notation we did not write explicitly the dependence of $\Lambda$ on $\lambda_\alpha$. The critical $\kappa_c$ is the value of $\kappa$ at which the dispersion relation has a double zero, namely the curve $\Lambda(\lambda_\alpha)$ touches the horizontal axis at $\lambda^{(c)}_\alpha$ and has a horizontal tangent at this point. Mathematically, this condition can be realized by imposing
\begin{equation*}
   \Lambda(\lambda^{(c)}_\alpha)=0\quad \text{and} \quad \frac{\partial\Lambda}{\partial \lambda_\alpha}\Big\rvert_{\lambda^{(c)}_\alpha}=0\, .
\end{equation*}
By using the definition~\eqref{eq:reldisp} and by performing the derivative, one can obtain the following two conditions:
\begin{equation} \label{simul1}
    \begin{array}{lcl}
       \text{det}(J_F(x^*))+\epsilon \lambda^{(c)}_\alpha (g_v+\kappa_c f_u)+\epsilon^2 (\lambda^{(c)}_\alpha)^2 \kappa_c&=&0,    \\
\epsilon (g_v+\kappa_c f_u)+2\epsilon^2\lambda^{(c)}_\alpha \kappa_c &=&0\,,
    \end{array}
\end{equation}
After some straightforward algebraic manipulations, we get
\begin{equation} \label{meanfield_sys}
    \kappa_c = \frac{f_ug_v-2f_vg_u+2\sqrt{f_vg_u(f_vg_u-f_ug_v)}}{f_u^2}\, .
\end{equation}
This analytical threshold, determined from ~\eqref{meanfield_sys}, determines when diffusion destabilizes the homogeneous coexistence state, as illustrated in Fig.~\ref{analytics}
and discussed in Sec.~\ref{3c}.

Let us observe that, as a byproduct, we have also been able to compute the critical eigenvalue $\lambda_\alpha^{(c)}$ that realizes the double zero of the dispersion relation. In conclusion, for all $\kappa < \kappa_c$, the dispersion relation is negative and patterns cannot emerge. On the other hand $\kappa > \kappa_c$ is a necessary condition to have patterns; indeed, because of finite size effects, one must ensure the existence of Laplace eigenvalues, $\lambda_\alpha$,  for which $\Lambda(\lambda_\alpha)>0$, the latter depends on the size of the network, on its topology but also on the model parameters (see SI.5, Fig. S6).
\subsection{Degree-based mean-field approximation} \label{meanfield_derive}
To analyze the heterogeneous spatial patterns observed numerically and analyzed in Sec.~\ref{2.4}, we employ a degree-based mean-field approximation applied to the studied networked system ~\eqref{network_system}. Using the definition of the Laplacian, the diffusive coupling term can be rewritten as $\sum_j L_{ij}u_j=\sum_j A_{ij}(u_j-u_i)$. Defining the average state of the neighbors of node $i$ as
$\bar{u}_i=\frac{1}{k_i}\sum_j A_{ij}u_j$,
the diffusion term becomes $\sum_j A_{ij}(u_j-u_i)=k_i(\bar{u}_i-u_i)$.
An analogous relation holds for the defector density $v_i$. Following the mean-field approximation commonly used for reaction–diffusion systems on complex networks \cite{nakao2009diffusion,nakao2010turing}, the neighbor average is approximated by a global degree-weighted mean field. We define
\begin{equation}
\langle u \rangle=\frac{\sum_j k_j u_j}{\sum_j k_j}, \qquad
\langle v \rangle=\frac{\sum_j k_j v_j}{\sum_j k_j},
\end{equation}
which represent the degree-weighted average densities of cooperators and defectors. This approximation accounts for the fact that highly connected nodes contribute more to a node's typical neighborhood. Under this assumption, $\bar{u}_i \approx \langle u \rangle,\; \bar{v}_i \approx \langle v\rangle $.
Substituting these expressions into the diffusion terms yields the reduced network dynamics as
\begin{equation}
\begin{aligned}
\dot{u}_i &= u_i[(1-u_i-v_i)(f_C+b)-d]+\epsilon k_i(\langle u\rangle-u_i), \\
\dot{v}_i &= v_i[(1-u_i-v_i)(f_D+b)-d]+\kappa\epsilon k_i(\langle v\rangle-v_i).
\end{aligned}
\end{equation}
Introducing the effective coupling parameter
$\beta_i=\epsilon k_i$, the reduced equations can be written as
\begin{equation}\label{meanfield_eq}
\begin{aligned}
\dot{u} &= u[(1-u-v)(f_C+b)-d]+\beta(\langle u\rangle-u), \\
\dot{v} &= v[(1-u-v)(f_D+b)-d]+\kappa\beta(\langle v\rangle-v),
\end{aligned}
\end{equation}
where we have omitted the index $i$, as all nodes follow the same equation. 
Within this approximation, nodes with different degrees experience different effective coupling strengths $\beta=\epsilon k$. Consequently, the heterogeneous network can be interpreted as an ensemble of identical eco-evolutionary systems evolving at different values of the control parameter $\beta$. The stationary states of the reduced system can therefore be analyzed using bifurcation methods, allowing the degree-dependent organization observed in the network simulations to be interpreted in terms of the bifurcation structure shown in Fig.~\ref{meanfield_cal}. 

Here, the global mean fields for cooperators, $\langle u\rangle$, and defectors, $\langle u\rangle$, are obtained self-consistently from the stationary patterned state of the full network dynamics \eqref{network_system}. The network system is first numerically integrated until a stationary configuration is reached. The global mean densities of cooperators and defectors are then computed as degree-weighted averages across all nodes and substituted into the reduced mean-field equations \eqref{meanfield_eq}. For fixed values of these mean fields, the stationary states of a node with a given degree are obtained from the steady-state conditions of the mean-field system. This yields degree-dependent branches of stationary solutions whose stability depends on the system parameters.
The stability of these solutions is determined from the Jacobian matrix of the mean-field system evaluated at the stationary states. The eigenvalues of the Jacobian characterize the stability of the solution branches, while saddle-node bifurcations occur when the determinant of the Jacobian vanishes, marking the appearance or disappearance of stationary states as the node degree varies. The resulting bifurcation structure provides the theoretical curves used to interpret the numerical results in Fig.~\ref{meanfield_cal}. {\color{black}We emphasize that the mean-field approximation captures the degree-dependent organization of the stationary patterns once the global mean densities are specified. However, the theory does not explain the increase in the average densities beyond the homogeneous equilibrium values, which is observed numerically.}

We further note that this mean-field approximation is not restricted to scale-free networks. A similar agreement is observed between the reduced equations and full network simulations for other network topologies, including small-world and random networks; these results are presented in the Supplementary Information (SI.6, Figs.~S7-S9). It is also important to note that the mean-field theory is generally not applicable for localized patterns below the instability threshold \cite{nakao2010turing}.}

\vspace{1cm}

\section*{Data availability} All data needed to evaluate the findings of the paper are available within the
paper itself. Additional data related to this paper are available from the
corresponding author upon reasonable request. The codes that were used for the findings of this study are openly available in
the GitHub repository \cite{web}.

\section*{Author Contributions}
M.S.A. designed the research; S.R. and M.S.A. performed the research; S.R., M.S.A., and T.C. contributed to the analytical results; S.R., M.S.A., T.C., M.P., and D.G. analyzed data; and S.R., M.S.A., T.C., M.P., and D.G. wrote the paper.

\section*{ACKNOWLEDGMENTS}M.P. was supported by the Slovenian Research and Innovation Agency (Grant No. P1-0403). M.S.A would like to thank Anusandhan National Research Foundation, Department of Science \& Technology, Government of India, for providing financial support in terms of a National Post Doctoral Fellowship (File Number: PDF/2025/003667). M.S.A would also like to thank the Indian Statistical Institute, Kolkata, and the Indian Institute of Technology Bombay for financial support during the preparation of the manuscript. The authors thank the reviewers for their insightful comments and constructive suggestions, which have significantly improved the manuscript.

\section*{author declaration} The authors declare no competing interests.


\begin{thebibliography}{47}%
	\makeatletter
	\providecommand \@ifxundefined [1]{%
		\@ifx{#1\undefined}
	}%
	\providecommand \@ifnum [1]{%
		\ifnum #1\expandafter \@firstoftwo
		\else \expandafter \@secondoftwo
		\fi
	}%
	\providecommand \@ifx [1]{%
		\ifx #1\expandafter \@firstoftwo
		\else \expandafter \@secondoftwo
		\fi
	}%
	\providecommand \natexlab [1]{#1}%
	\providecommand \enquote  [1]{``#1''}%
	\providecommand \bibnamefont  [1]{#1}%
	\providecommand \bibfnamefont [1]{#1}%
	\providecommand \citenamefont [1]{#1}%
	\providecommand \href@noop [0]{\@secondoftwo}%
	\providecommand \href [0]{\begingroup \@sanitize@url \@href}%
	\providecommand \@href[1]{\@@startlink{#1}\@@href}%
	\providecommand \@@href[1]{\endgroup#1\@@endlink}%
	\providecommand \@sanitize@url [0]{\catcode `\\12\catcode `\$12\catcode
		`\&12\catcode `\#12\catcode `\^12\catcode `\_12\catcode `\%12\relax}%
	\providecommand \@@startlink[1]{}%
	\providecommand \@@endlink[0]{}%
	\providecommand \url  [0]{\begingroup\@sanitize@url \@url }%
	\providecommand \@url [1]{\endgroup\@href {#1}{\urlprefix }}%
	\providecommand \urlprefix  [0]{URL }%
	\providecommand \Eprint [0]{\href }%
	\providecommand \doibase [0]{https://doi.org/}%
	\providecommand \selectlanguage [0]{\@gobble}%
	\providecommand \bibinfo  [0]{\@secondoftwo}%
	\providecommand \bibfield  [0]{\@secondoftwo}%
	\providecommand \translation [1]{[#1]}%
	\providecommand \BibitemOpen [0]{}%
	\providecommand \bibitemStop [0]{}%
	\providecommand \bibitemNoStop [0]{.\EOS\space}%
	\providecommand \EOS [0]{\spacefactor3000\relax}%
	\providecommand \BibitemShut  [1]{\csname bibitem#1\endcsname}%
	\let\auto@bib@innerbib\@empty
	\bibitem [{\citenamefont {Szolnoki}\ and\ \citenamefont
		{Perc}(2010)}]{szolnoki2010reward}%
	\BibitemOpen
	\bibfield  {author} {\bibinfo {author} {\bibfnamefont {A.}~\bibnamefont
			{Szolnoki}}\ and\ \bibinfo {author} {\bibfnamefont {M.}~\bibnamefont
			{Perc}},\ }\href@noop {} {\bibfield  {journal} {\bibinfo  {journal}
			{Europhysics Letters}\ }\textbf {\bibinfo {volume} {92}},\ \bibinfo {pages}
		{38003} (\bibinfo {year} {2010})}\BibitemShut {NoStop}%
	\bibitem [{\citenamefont {Balduzzi}\ \emph {et~al.}(2018)\citenamefont
		{Balduzzi}, \citenamefont {Racaniere}, \citenamefont {Martens}, \citenamefont
		{Foerster}, \citenamefont {Tuyls},\ and\ \citenamefont
		{Graepel}}]{balduzzi2018mechanics}%
	\BibitemOpen
	\bibfield  {author} {\bibinfo {author} {\bibfnamefont {D.}~\bibnamefont
			{Balduzzi}}, \bibinfo {author} {\bibfnamefont {S.}~\bibnamefont {Racaniere}},
		\bibinfo {author} {\bibfnamefont {J.}~\bibnamefont {Martens}}, \bibinfo
		{author} {\bibfnamefont {J.}~\bibnamefont {Foerster}}, \bibinfo {author}
		{\bibfnamefont {K.}~\bibnamefont {Tuyls}},\ and\ \bibinfo {author}
		{\bibfnamefont {T.}~\bibnamefont {Graepel}},\ }in\ \href@noop {} {\emph
		{\bibinfo {booktitle} {International Conference on Machine Learning}}}\
	(\bibinfo {organization} {PMLR},\ \bibinfo {year} {2018})\ pp.\ \bibinfo
	{pages} {354--363}\BibitemShut {NoStop}%
	\bibitem [{\citenamefont {Nowak}(2006{\natexlab{a}})}]{nowak2006five}%
	\BibitemOpen
	\bibfield  {author} {\bibinfo {author} {\bibfnamefont {M.~A.}\ \bibnamefont
			{Nowak}},\ }\href@noop {} {\bibfield  {journal} {\bibinfo  {journal}
			{Science}\ }\textbf {\bibinfo {volume} {314}},\ \bibinfo {pages} {1560}
		(\bibinfo {year} {2006}{\natexlab{a}})}\BibitemShut {NoStop}%
	\bibitem [{\citenamefont {Szab{\'o}}\ and\ \citenamefont
		{Hauert}(2002)}]{szabo2002phase}%
	\BibitemOpen
	\bibfield  {author} {\bibinfo {author} {\bibfnamefont {G.}~\bibnamefont
			{Szab{\'o}}}\ and\ \bibinfo {author} {\bibfnamefont {C.}~\bibnamefont
			{Hauert}},\ }\href@noop {} {\bibfield  {journal} {\bibinfo  {journal}
			{Physical Review Letters}\ }\textbf {\bibinfo {volume} {89}},\ \bibinfo
		{pages} {118101} (\bibinfo {year} {2002})}\BibitemShut {NoStop}%
	\bibitem [{\citenamefont {Hardin}(2018)}]{hardin2018tragedy}%
	\BibitemOpen
	\bibfield  {author} {\bibinfo {author} {\bibfnamefont {G.}~\bibnamefont
			{Hardin}},\ }in\ \href@noop {} {\emph {\bibinfo {booktitle} {Classic Papers
				in Natural Resource Economics Revisited}}}\ (\bibinfo  {publisher}
	{Routledge},\ \bibinfo {year} {2018})\ pp.\ \bibinfo {pages}
	{145--156}\BibitemShut {NoStop}%
	\bibitem [{\citenamefont {Ostrom}(2008)}]{ostrom2008tragedy}%
	\BibitemOpen
	\bibfield  {author} {\bibinfo {author} {\bibfnamefont {E.}~\bibnamefont
			{Ostrom}},\ }\href@noop {} {\bibfield  {journal} {\bibinfo  {journal} {The
				New Palgrave Dictionary of Economics}\ }\textbf {\bibinfo {volume} {2}},\
		\bibinfo {pages} {1} (\bibinfo {year} {2008})}\BibitemShut {NoStop}%
	\bibitem [{\citenamefont {Sohel~Mondal}\ \emph {et~al.}(2024)\citenamefont
		{Sohel~Mondal}, \citenamefont {Ray},\ and\ \citenamefont
		{Chakraborty}}]{sohel2024hypochaos}%
	\BibitemOpen
	\bibfield  {author} {\bibinfo {author} {\bibfnamefont {S.}~\bibnamefont
			{Sohel~Mondal}}, \bibinfo {author} {\bibfnamefont {A.}~\bibnamefont {Ray}},\
		and\ \bibinfo {author} {\bibfnamefont {S.}~\bibnamefont {Chakraborty}},\
	}\href@noop {} {\bibfield  {journal} {\bibinfo  {journal} {Chaos: An
				Interdisciplinary Journal of Nonlinear Science}\ }\textbf {\bibinfo {volume}
			{34}},\ \bibinfo {pages} {023122} (\bibinfo {year} {2024})}\BibitemShut
	{NoStop}%
	\bibitem [{\citenamefont {Nowak}(2006{\natexlab{b}})}]{nowak2006evolutionary}%
	\BibitemOpen
	\bibfield  {author} {\bibinfo {author} {\bibfnamefont {M.~A.}\ \bibnamefont
			{Nowak}},\ }in\ \href@noop {} {\emph {\bibinfo {booktitle} {Proceedings of
				the International Congress of Mathematicians}}},\ Vol.~\bibinfo {volume} {3}\
	(\bibinfo {year} {2006})\ pp.\ \bibinfo {pages} {1523--40}\BibitemShut
	{NoStop}%
	\bibitem [{\citenamefont {Nadell}\ \emph {et~al.}(2016)\citenamefont {Nadell},
		\citenamefont {Drescher},\ and\ \citenamefont {Foster}}]{nadell2016spatial}%
	\BibitemOpen
	\bibfield  {author} {\bibinfo {author} {\bibfnamefont {C.~D.}\ \bibnamefont
			{Nadell}}, \bibinfo {author} {\bibfnamefont {K.}~\bibnamefont {Drescher}},\
		and\ \bibinfo {author} {\bibfnamefont {K.~R.}\ \bibnamefont {Foster}},\
	}\href@noop {} {\bibfield  {journal} {\bibinfo  {journal} {Nature Reviews
				Microbiology}\ }\textbf {\bibinfo {volume} {14}},\ \bibinfo {pages} {589}
		(\bibinfo {year} {2016})}\BibitemShut {NoStop}%
	\bibitem [{\citenamefont {Cremer}\ \emph {et~al.}(2019)\citenamefont {Cremer},
		\citenamefont {Honda}, \citenamefont {Tang}, \citenamefont {Wong-Ng},
		\citenamefont {Vergassola},\ and\ \citenamefont
		{Hwa}}]{cremer2019chemotaxis}%
	\BibitemOpen
	\bibfield  {author} {\bibinfo {author} {\bibfnamefont {J.}~\bibnamefont
			{Cremer}}, \bibinfo {author} {\bibfnamefont {T.}~\bibnamefont {Honda}},
		\bibinfo {author} {\bibfnamefont {Y.}~\bibnamefont {Tang}}, \bibinfo {author}
		{\bibfnamefont {J.}~\bibnamefont {Wong-Ng}}, \bibinfo {author} {\bibfnamefont
			{M.}~\bibnamefont {Vergassola}},\ and\ \bibinfo {author} {\bibfnamefont
			{T.}~\bibnamefont {Hwa}},\ }\href@noop {} {\bibfield  {journal} {\bibinfo
			{journal} {Nature}\ }\textbf {\bibinfo {volume} {575}},\ \bibinfo {pages}
		{658} (\bibinfo {year} {2019})}\BibitemShut {NoStop}%
	\bibitem [{\citenamefont {Rand}\ \emph {et~al.}(2014)\citenamefont {Rand},
		\citenamefont {Peysakhovich}, \citenamefont {Kraft-Todd}, \citenamefont
		{Newman}, \citenamefont {Wurzbacher}, \citenamefont {Nowak},\ and\
		\citenamefont {Greene}}]{rand2014social}%
	\BibitemOpen
	\bibfield  {author} {\bibinfo {author} {\bibfnamefont {D.~G.}\ \bibnamefont
			{Rand}}, \bibinfo {author} {\bibfnamefont {A.}~\bibnamefont {Peysakhovich}},
		\bibinfo {author} {\bibfnamefont {G.~T.}\ \bibnamefont {Kraft-Todd}},
		\bibinfo {author} {\bibfnamefont {G.~E.}\ \bibnamefont {Newman}}, \bibinfo
		{author} {\bibfnamefont {O.}~\bibnamefont {Wurzbacher}}, \bibinfo {author}
		{\bibfnamefont {M.~A.}\ \bibnamefont {Nowak}},\ and\ \bibinfo {author}
		{\bibfnamefont {J.~D.}\ \bibnamefont {Greene}},\ }\href@noop {} {\bibfield
		{journal} {\bibinfo  {journal} {Nature Communications}\ }\textbf {\bibinfo
			{volume} {5}},\ \bibinfo {pages} {3677} (\bibinfo {year} {2014})}\BibitemShut
	{NoStop}%
	\bibitem [{\citenamefont {Roy}\ \emph {et~al.}(2023)\citenamefont {Roy},
		\citenamefont {Nag~Chowdhury}, \citenamefont {Kundu}, \citenamefont {Sar},
		\citenamefont {Banerjee}, \citenamefont {Rakshit}, \citenamefont {Mali},
		\citenamefont {Perc},\ and\ \citenamefont {Ghosh}}]{roy2023time}%
	\BibitemOpen
	\bibfield  {author} {\bibinfo {author} {\bibfnamefont {S.}~\bibnamefont
			{Roy}}, \bibinfo {author} {\bibfnamefont {S.}~\bibnamefont {Nag~Chowdhury}},
		\bibinfo {author} {\bibfnamefont {S.}~\bibnamefont {Kundu}}, \bibinfo
		{author} {\bibfnamefont {G.~K.}\ \bibnamefont {Sar}}, \bibinfo {author}
		{\bibfnamefont {J.}~\bibnamefont {Banerjee}}, \bibinfo {author}
		{\bibfnamefont {B.}~\bibnamefont {Rakshit}}, \bibinfo {author} {\bibfnamefont
			{P.~C.}\ \bibnamefont {Mali}}, \bibinfo {author} {\bibfnamefont
			{M.}~\bibnamefont {Perc}},\ and\ \bibinfo {author} {\bibfnamefont
			{D.}~\bibnamefont {Ghosh}},\ }\href@noop {} {\bibfield  {journal} {\bibinfo
			{journal} {Scientific Reports}\ }\textbf {\bibinfo {volume} {13}},\ \bibinfo
		{pages} {14331} (\bibinfo {year} {2023})}\BibitemShut {NoStop}%
	\bibitem [{\citenamefont {Betz}\ \emph {et~al.}(2024)\citenamefont {Betz},
		\citenamefont {Fu},\ and\ \citenamefont {Masuda}}]{betz2024evolutionary}%
	\BibitemOpen
	\bibfield  {author} {\bibinfo {author} {\bibfnamefont {K.}~\bibnamefont
			{Betz}}, \bibinfo {author} {\bibfnamefont {F.}~\bibnamefont {Fu}},\ and\
		\bibinfo {author} {\bibfnamefont {N.}~\bibnamefont {Masuda}},\ }\href@noop {}
	{\bibfield  {journal} {\bibinfo  {journal} {Bulletin of Mathematical
				Biology}\ }\textbf {\bibinfo {volume} {86}},\ \bibinfo {pages} {84} (\bibinfo
		{year} {2024})}\BibitemShut {NoStop}%
	\bibitem [{\citenamefont {Chowdhury}\ \emph {et~al.}(2021)\citenamefont
		{Chowdhury}, \citenamefont {Kundu}, \citenamefont {Banerjee}, \citenamefont
		{Perc},\ and\ \citenamefont {Ghosh}}]{chowdhury2021eco}%
	\BibitemOpen
	\bibfield  {author} {\bibinfo {author} {\bibfnamefont {S.~N.}\ \bibnamefont
			{Chowdhury}}, \bibinfo {author} {\bibfnamefont {S.}~\bibnamefont {Kundu}},
		\bibinfo {author} {\bibfnamefont {J.}~\bibnamefont {Banerjee}}, \bibinfo
		{author} {\bibfnamefont {M.}~\bibnamefont {Perc}},\ and\ \bibinfo {author}
		{\bibfnamefont {D.}~\bibnamefont {Ghosh}},\ }\href@noop {} {\bibfield
		{journal} {\bibinfo  {journal} {Journal of Theoretical Biology}\ }\textbf
		{\bibinfo {volume} {518}},\ \bibinfo {pages} {110606} (\bibinfo {year}
		{2021})}\BibitemShut {NoStop}%
	\bibitem [{\citenamefont {Wakano}\ \emph {et~al.}(2009)\citenamefont {Wakano},
		\citenamefont {Nowak},\ and\ \citenamefont {Hauert}}]{wakano2009spatial}%
	\BibitemOpen
	\bibfield  {author} {\bibinfo {author} {\bibfnamefont {J.~Y.}\ \bibnamefont
			{Wakano}}, \bibinfo {author} {\bibfnamefont {M.~A.}\ \bibnamefont {Nowak}},\
		and\ \bibinfo {author} {\bibfnamefont {C.}~\bibnamefont {Hauert}},\
	}\href@noop {} {\bibfield  {journal} {\bibinfo  {journal} {Proceedings of the
				National Academy of Sciences}\ }\textbf {\bibinfo {volume} {106}},\ \bibinfo
		{pages} {7910} (\bibinfo {year} {2009})}\BibitemShut {NoStop}%
	\bibitem [{\citenamefont {Wakano}\ and\ \citenamefont
		{Hauert}(2011)}]{wakano2011pattern}%
	\BibitemOpen
	\bibfield  {author} {\bibinfo {author} {\bibfnamefont {J.~Y.}\ \bibnamefont
			{Wakano}}\ and\ \bibinfo {author} {\bibfnamefont {C.}~\bibnamefont
			{Hauert}},\ }\href@noop {} {\bibfield  {journal} {\bibinfo  {journal}
			{Journal of Theoretical Biology}\ }\textbf {\bibinfo {volume} {268}},\
		\bibinfo {pages} {30} (\bibinfo {year} {2011})}\BibitemShut {NoStop}%
	\bibitem [{\citenamefont {Hernandez}\ \emph {et~al.}(2021)\citenamefont
		{Hernandez}, \citenamefont {David}, \citenamefont {Menges}, \citenamefont
		{Searcy},\ and\ \citenamefont {Afkhami}}]{hernandez2021environmental}%
	\BibitemOpen
	\bibfield  {author} {\bibinfo {author} {\bibfnamefont {D.~J.}\ \bibnamefont
			{Hernandez}}, \bibinfo {author} {\bibfnamefont {A.~S.}\ \bibnamefont
			{David}}, \bibinfo {author} {\bibfnamefont {E.~S.}\ \bibnamefont {Menges}},
		\bibinfo {author} {\bibfnamefont {C.~A.}\ \bibnamefont {Searcy}},\ and\
		\bibinfo {author} {\bibfnamefont {M.~E.}\ \bibnamefont {Afkhami}},\
	}\href@noop {} {\bibfield  {journal} {\bibinfo  {journal} {The ISME journal}\
		}\textbf {\bibinfo {volume} {15}},\ \bibinfo {pages} {1722} (\bibinfo {year}
		{2021})}\BibitemShut {NoStop}%
	\bibitem [{\citenamefont {Pan}\ \emph {et~al.}(2023)\citenamefont {Pan},
		\citenamefont {Zhang}, \citenamefont {Han},\ and\ \citenamefont
		{Huang}}]{pan2023heterogeneous}%
	\BibitemOpen
	\bibfield  {author} {\bibinfo {author} {\bibfnamefont {J.}~\bibnamefont
			{Pan}}, \bibinfo {author} {\bibfnamefont {L.}~\bibnamefont {Zhang}}, \bibinfo
		{author} {\bibfnamefont {W.}~\bibnamefont {Han}},\ and\ \bibinfo {author}
		{\bibfnamefont {C.}~\bibnamefont {Huang}},\ }\href@noop {} {\bibfield
		{journal} {\bibinfo  {journal} {Physica A: Statistical Mechanics and its
				Applications}\ }\textbf {\bibinfo {volume} {609}},\ \bibinfo {pages} {128400}
		(\bibinfo {year} {2023})}\BibitemShut {NoStop}%
	\bibitem [{\citenamefont {Pires}\ and\ \citenamefont
		{Broom}(2024)}]{pires2024rules}%
	\BibitemOpen
	\bibfield  {author} {\bibinfo {author} {\bibfnamefont {D.~L.}\ \bibnamefont
			{Pires}}\ and\ \bibinfo {author} {\bibfnamefont {M.}~\bibnamefont {Broom}},\
	}\href@noop {} {\bibfield  {journal} {\bibinfo  {journal} {PLOS Computational
				Biology}\ }\textbf {\bibinfo {volume} {20}},\ \bibinfo {pages} {e1012388}
		(\bibinfo {year} {2024})}\BibitemShut {NoStop}%
	\bibitem [{\citenamefont {Barab{\'a}si}\ and\ \citenamefont
		{Bonabeau}(2003)}]{barabasi2003scale}%
	\BibitemOpen
	\bibfield  {author} {\bibinfo {author} {\bibfnamefont {A.-L.}\ \bibnamefont
			{Barab{\'a}si}}\ and\ \bibinfo {author} {\bibfnamefont {E.}~\bibnamefont
			{Bonabeau}},\ }\href@noop {} {\bibfield  {journal} {\bibinfo  {journal}
			{Scientific American}\ }\textbf {\bibinfo {volume} {288}},\ \bibinfo {pages}
		{60} (\bibinfo {year} {2003})}\BibitemShut {NoStop}%
	\bibitem [{\citenamefont {Watts}\ and\ \citenamefont
		{Strogatz}(1998)}]{watts1998collective}%
	\BibitemOpen
	\bibfield  {author} {\bibinfo {author} {\bibfnamefont {D.~J.}\ \bibnamefont
			{Watts}}\ and\ \bibinfo {author} {\bibfnamefont {S.~H.}\ \bibnamefont
			{Strogatz}},\ }\href@noop {} {\bibfield  {journal} {\bibinfo  {journal}
			{Nature}\ }\textbf {\bibinfo {volume} {393}},\ \bibinfo {pages} {440}
		(\bibinfo {year} {1998})}\BibitemShut {NoStop}%
	\bibitem [{\citenamefont {Andreas}\ \emph {et~al.}(2016)\citenamefont
		{Andreas}, \citenamefont {Rohrbach}, \citenamefont {Darrell},\ and\
		\citenamefont {Klein}}]{andreas2016neural}%
	\BibitemOpen
	\bibfield  {author} {\bibinfo {author} {\bibfnamefont {J.}~\bibnamefont
			{Andreas}}, \bibinfo {author} {\bibfnamefont {M.}~\bibnamefont {Rohrbach}},
		\bibinfo {author} {\bibfnamefont {T.}~\bibnamefont {Darrell}},\ and\ \bibinfo
		{author} {\bibfnamefont {D.}~\bibnamefont {Klein}},\ }in\ \href@noop {}
	{\emph {\bibinfo {booktitle} {Proceedings of the IEEE conference on computer
				vision and pattern recognition}}}\ (\bibinfo {year} {2016})\ pp.\ \bibinfo
	{pages} {39--48}\BibitemShut {NoStop}%
	\bibitem [{\citenamefont {Santos}\ and\ \citenamefont
		{Pacheco}(2005)}]{santos2005scale}%
	\BibitemOpen
	\bibfield  {author} {\bibinfo {author} {\bibfnamefont {F.~C.}\ \bibnamefont
			{Santos}}\ and\ \bibinfo {author} {\bibfnamefont {J.~M.}\ \bibnamefont
			{Pacheco}},\ }\href@noop {} {\bibfield  {journal} {\bibinfo  {journal}
			{Physical Review Letters}\ }\textbf {\bibinfo {volume} {95}},\ \bibinfo
		{pages} {098104} (\bibinfo {year} {2005})}\BibitemShut {NoStop}%
	\bibitem [{\citenamefont {Nakao}\ and\ \citenamefont
		{Mikhailov}(2010)}]{nakao2010turing}%
	\BibitemOpen
	\bibfield  {author} {\bibinfo {author} {\bibfnamefont {H.}~\bibnamefont
			{Nakao}}\ and\ \bibinfo {author} {\bibfnamefont {A.~S.}\ \bibnamefont
			{Mikhailov}},\ }\href@noop {} {\bibfield  {journal} {\bibinfo  {journal}
			{Nature Physics}\ }\textbf {\bibinfo {volume} {6}},\ \bibinfo {pages} {544}
		(\bibinfo {year} {2010})}\BibitemShut {NoStop}%
	\bibitem [{\citenamefont {Gao}\ \emph {et~al.}(2023)\citenamefont {Gao},
		\citenamefont {Chang}, \citenamefont {Perc},\ and\ \citenamefont
		{Wang}}]{gao2023turing}%
	\BibitemOpen
	\bibfield  {author} {\bibinfo {author} {\bibfnamefont {S.}~\bibnamefont
			{Gao}}, \bibinfo {author} {\bibfnamefont {L.}~\bibnamefont {Chang}}, \bibinfo
		{author} {\bibfnamefont {M.}~\bibnamefont {Perc}},\ and\ \bibinfo {author}
		{\bibfnamefont {Z.}~\bibnamefont {Wang}},\ }\href@noop {} {\bibfield
		{journal} {\bibinfo  {journal} {Physical Review E}\ }\textbf {\bibinfo
			{volume} {107}},\ \bibinfo {pages} {014216} (\bibinfo {year}
		{2023})}\BibitemShut {NoStop}%
	\bibitem [{\citenamefont {Stanley}\ \emph {et~al.}(1995)\citenamefont
		{Stanley}, \citenamefont {Ashlock},\ and\ \citenamefont
		{Smucker}}]{stanley1995iterated}%
	\BibitemOpen
	\bibfield  {author} {\bibinfo {author} {\bibfnamefont {E.~A.}\ \bibnamefont
			{Stanley}}, \bibinfo {author} {\bibfnamefont {D.}~\bibnamefont {Ashlock}},\
		and\ \bibinfo {author} {\bibfnamefont {M.~D.}\ \bibnamefont {Smucker}},\ }in\
	\href@noop {} {\emph {\bibinfo {booktitle} {European Conference on Artificial
				Life}}}\ (\bibinfo {organization} {Springer},\ \bibinfo {year} {1995})\ pp.\
	\bibinfo {pages} {490--502}\BibitemShut {NoStop}%
	\bibitem [{\citenamefont {Ashlock}\ \emph {et~al.}(1996)\citenamefont
		{Ashlock}, \citenamefont {Smucker}, \citenamefont {Stanley},\ and\
		\citenamefont {Tesfatsion}}]{ashlock1996preferential}%
	\BibitemOpen
	\bibfield  {author} {\bibinfo {author} {\bibfnamefont {D.}~\bibnamefont
			{Ashlock}}, \bibinfo {author} {\bibfnamefont {M.~D.}\ \bibnamefont
			{Smucker}}, \bibinfo {author} {\bibfnamefont {E.~A.}\ \bibnamefont
			{Stanley}},\ and\ \bibinfo {author} {\bibfnamefont {L.}~\bibnamefont
			{Tesfatsion}},\ }\href@noop {} {\bibfield  {journal} {\bibinfo  {journal}
			{BioSystems}\ }\textbf {\bibinfo {volume} {37}},\ \bibinfo {pages} {99}
		(\bibinfo {year} {1996})}\BibitemShut {NoStop}%
	\bibitem [{\citenamefont {Hu}\ and\ \citenamefont
		{Chen}(2024)}]{hu2024collective}%
	\BibitemOpen
	\bibfield  {author} {\bibinfo {author} {\bibfnamefont {M.}~\bibnamefont
			{Hu}}\ and\ \bibinfo {author} {\bibfnamefont {W.}~\bibnamefont {Chen}},\
	}\href@noop {} {\bibfield  {journal} {\bibinfo  {journal} {Chaos, Solitons \&
				Fractals}\ }\textbf {\bibinfo {volume} {184}},\ \bibinfo {pages} {115058}
		(\bibinfo {year} {2024})}\BibitemShut {NoStop}%
	\bibitem [{\citenamefont {Zhang}\ \emph {et~al.}(2025)\citenamefont {Zhang},
		\citenamefont {Yao}, \citenamefont {Zeng}, \citenamefont {Feng},\ and\
		\citenamefont {Chica}}]{zhang2025evolution}%
	\BibitemOpen
	\bibfield  {author} {\bibinfo {author} {\bibfnamefont {G.}~\bibnamefont
			{Zhang}}, \bibinfo {author} {\bibfnamefont {Y.}~\bibnamefont {Yao}}, \bibinfo
		{author} {\bibfnamefont {Z.}~\bibnamefont {Zeng}}, \bibinfo {author}
		{\bibfnamefont {M.}~\bibnamefont {Feng}},\ and\ \bibinfo {author}
		{\bibfnamefont {M.}~\bibnamefont {Chica}},\ }\href@noop {} {\bibfield
		{journal} {\bibinfo  {journal} {Chaos: An Interdisciplinary Journal of
				Nonlinear Science}\ }\textbf {\bibinfo {volume} {35}},\ \bibinfo {pages}
		{013104} (\bibinfo {year} {2025})}\BibitemShut {NoStop}%
	\bibitem [{\citenamefont {Cheng}\ \emph {et~al.}(2024)\citenamefont {Cheng},
		\citenamefont {Sysoeva}, \citenamefont {Wang}, \citenamefont {Yuan},
		\citenamefont {Zhang},\ and\ \citenamefont {Meng}}]{cheng2024evolution}%
	\BibitemOpen
	\bibfield  {author} {\bibinfo {author} {\bibfnamefont {H.}~\bibnamefont
			{Cheng}}, \bibinfo {author} {\bibfnamefont {L.}~\bibnamefont {Sysoeva}},
		\bibinfo {author} {\bibfnamefont {H.}~\bibnamefont {Wang}}, \bibinfo {author}
		{\bibfnamefont {H.}~\bibnamefont {Yuan}}, \bibinfo {author} {\bibfnamefont
			{T.}~\bibnamefont {Zhang}},\ and\ \bibinfo {author} {\bibfnamefont
			{X.}~\bibnamefont {Meng}},\ }\href@noop {} {\bibfield  {journal} {\bibinfo
			{journal} {Bulletin of Mathematical Biology}\ }\textbf {\bibinfo {volume}
			{86}},\ \bibinfo {pages} {67} (\bibinfo {year} {2024})}\BibitemShut {NoStop}%
	\bibitem [{\citenamefont {Turing}(1952)}]{Turing1952}%
	\BibitemOpen
	\bibfield  {author} {\bibinfo {author} {\bibfnamefont {A.}~\bibnamefont
			{Turing}},\ }\href@noop {} {\bibfield  {journal} {\bibinfo  {journal} {Phil.
				Trans. R. Soc. Lond. B}\ }\textbf {\bibinfo {volume} {237}},\ \bibinfo
		{pages} {37} (\bibinfo {year} {1952})}\BibitemShut {NoStop}%
	\bibitem [{\citenamefont {Gokhale}\ and\ \citenamefont
		{Traulsen}(2010)}]{gokhale2010evolutionary}%
	\BibitemOpen
	\bibfield  {author} {\bibinfo {author} {\bibfnamefont {C.~S.}\ \bibnamefont
			{Gokhale}}\ and\ \bibinfo {author} {\bibfnamefont {A.}~\bibnamefont
			{Traulsen}},\ }\href@noop {} {\bibfield  {journal} {\bibinfo  {journal}
			{Proceedings of the National Academy of Sciences}\ }\textbf {\bibinfo
			{volume} {107}},\ \bibinfo {pages} {5500} (\bibinfo {year}
		{2010})}\BibitemShut {NoStop}%
	\bibitem [{\citenamefont {Gokhale}\ and\ \citenamefont
		{Hauert}(2016)}]{gokhale2016eco}%
	\BibitemOpen
	\bibfield  {author} {\bibinfo {author} {\bibfnamefont {C.~S.}\ \bibnamefont
			{Gokhale}}\ and\ \bibinfo {author} {\bibfnamefont {C.}~\bibnamefont
			{Hauert}},\ }\href@noop {} {\bibfield  {journal} {\bibinfo  {journal}
			{Theoretical Population Biology}\ }\textbf {\bibinfo {volume} {111}},\
		\bibinfo {pages} {28} (\bibinfo {year} {2016})}\BibitemShut {NoStop}%
	\bibitem [{\citenamefont {Merris}(1994)}]{merris1994laplacian}%
	\BibitemOpen
	\bibfield  {author} {\bibinfo {author} {\bibfnamefont {R.}~\bibnamefont
			{Merris}},\ }\href@noop {} {\bibfield  {journal} {\bibinfo  {journal} {Linear
				Algebra and its Applications}\ }\textbf {\bibinfo {volume} {197}},\ \bibinfo
		{pages} {143} (\bibinfo {year} {1994})}\BibitemShut {NoStop}%
	\bibitem [{\citenamefont {Asllani}\ \emph {et~al.}(2014)\citenamefont
		{Asllani}, \citenamefont {Busiello}, \citenamefont {Carletti}, \citenamefont
		{Fanelli},\ and\ \citenamefont {Planchon}}]{AsllaniPRE2014}%
	\BibitemOpen
	\bibfield  {author} {\bibinfo {author} {\bibfnamefont {M.}~\bibnamefont
			{Asllani}}, \bibinfo {author} {\bibfnamefont {D.~M.}\ \bibnamefont
			{Busiello}}, \bibinfo {author} {\bibfnamefont {T.}~\bibnamefont {Carletti}},
		\bibinfo {author} {\bibfnamefont {D.}~\bibnamefont {Fanelli}},\ and\ \bibinfo
		{author} {\bibfnamefont {G.}~\bibnamefont {Planchon}},\ }\href@noop {}
	{\bibfield  {journal} {\bibinfo  {journal} {Physical Review E}\ }\textbf
		{\bibinfo {volume} {90}},\ \bibinfo {pages} {042814} (\bibinfo {year}
		{2014})}\BibitemShut {NoStop}%
	\bibitem [{\citenamefont {Muolo}\ \emph {et~al.}(2024)\citenamefont {Muolo},
		\citenamefont {Giambagli}, \citenamefont {Nakao}, \citenamefont {Fanelli},\
		and\ \citenamefont {Carletti}}]{MuoloPRSA2024}%
	\BibitemOpen
	\bibfield  {author} {\bibinfo {author} {\bibfnamefont {R.}~\bibnamefont
			{Muolo}}, \bibinfo {author} {\bibfnamefont {L.}~\bibnamefont {Giambagli}},
		\bibinfo {author} {\bibfnamefont {H.}~\bibnamefont {Nakao}}, \bibinfo
		{author} {\bibfnamefont {D.}~\bibnamefont {Fanelli}},\ and\ \bibinfo {author}
		{\bibfnamefont {T.}~\bibnamefont {Carletti}},\ }\href@noop {} {\bibfield
		{journal} {\bibinfo  {journal} {Proc. R. Soc. A.}\ }\textbf {\bibinfo
			{volume} {480}},\ \bibinfo {pages} {20240235} (\bibinfo {year}
		{2024})}\BibitemShut {NoStop}%
	\bibitem [{\citenamefont {Hauert}\ \emph {et~al.}(2006)\citenamefont {Hauert},
		\citenamefont {Holmes},\ and\ \citenamefont
		{Doebeli}}]{hauert2006evolutionary}%
	\BibitemOpen
	\bibfield  {author} {\bibinfo {author} {\bibfnamefont {C.}~\bibnamefont
			{Hauert}}, \bibinfo {author} {\bibfnamefont {M.}~\bibnamefont {Holmes}},\
		and\ \bibinfo {author} {\bibfnamefont {M.}~\bibnamefont {Doebeli}},\
	}\href@noop {} {\bibfield  {journal} {\bibinfo  {journal} {Proceedings of the
				Royal Society B: Biological Sciences}\ }\textbf {\bibinfo {volume} {273}},\
		\bibinfo {pages} {2565} (\bibinfo {year} {2006})}\BibitemShut {NoStop}%
	\bibitem [{\citenamefont {Hauert}\ \emph {et~al.}(2008)\citenamefont {Hauert},
		\citenamefont {Wakano},\ and\ \citenamefont
		{Doebeli}}]{hauert2008ecological}%
	\BibitemOpen
	\bibfield  {author} {\bibinfo {author} {\bibfnamefont {C.}~\bibnamefont
			{Hauert}}, \bibinfo {author} {\bibfnamefont {J.~Y.}\ \bibnamefont {Wakano}},\
		and\ \bibinfo {author} {\bibfnamefont {M.}~\bibnamefont {Doebeli}},\
	}\href@noop {} {\bibfield  {journal} {\bibinfo  {journal} {Theoretical
				Population Biology}\ }\textbf {\bibinfo {volume} {73}},\ \bibinfo {pages}
		{257} (\bibinfo {year} {2008})}\BibitemShut {NoStop}%
	\bibitem [{\citenamefont {May}(1972)}]{RobertMay1972}%
	\BibitemOpen
	\bibfield  {author} {\bibinfo {author} {\bibfnamefont {R.}~\bibnamefont
			{May}},\ }\href@noop {} {\bibfield  {journal} {\bibinfo  {journal} {Nature}\
		}\textbf {\bibinfo {volume} {238}},\ \bibinfo {pages} {413} (\bibinfo {year}
		{1972})}\BibitemShut {NoStop}%
	\bibitem [{\citenamefont {Barrat}\ \emph {et~al.}(2008)\citenamefont {Barrat},
		\citenamefont {Barthelemy},\ and\ \citenamefont
		{Vespignani}}]{barrat2008dynamical}%
	\BibitemOpen
	\bibfield  {author} {\bibinfo {author} {\bibfnamefont {A.}~\bibnamefont
			{Barrat}}, \bibinfo {author} {\bibfnamefont {M.}~\bibnamefont {Barthelemy}},\
		and\ \bibinfo {author} {\bibfnamefont {A.}~\bibnamefont {Vespignani}},\
	}\href@noop {} {\emph {\bibinfo {title} {Dynamical processes on complex
				networks}}}\ (\bibinfo  {publisher} {Cambridge university press},\ \bibinfo
	{year} {2008})\BibitemShut {NoStop}%
	\bibitem [{\citenamefont {Albert}\ and\ \citenamefont
		{Barab{\'a}si}(2002)}]{BarabasiAlbert2002}%
	\BibitemOpen
	\bibfield  {author} {\bibinfo {author} {\bibfnamefont {R.}~\bibnamefont
			{Albert}}\ and\ \bibinfo {author} {\bibfnamefont {A.-L.}\ \bibnamefont
			{Barab{\'a}si}},\ }\href@noop {} {\bibfield  {journal} {\bibinfo  {journal}
			{Reviews of Modern Physics}\ }\textbf {\bibinfo {volume} {74}},\ \bibinfo
		{pages} {47} (\bibinfo {year} {2002})}\BibitemShut {NoStop}%
	\bibitem [{\citenamefont {Nakao}\ and\ \citenamefont
		{Mikhailov}(2009)}]{nakao2009diffusion}%
	\BibitemOpen
	\bibfield  {author} {\bibinfo {author} {\bibfnamefont {H.}~\bibnamefont
			{Nakao}}\ and\ \bibinfo {author} {\bibfnamefont {A.~S.}\ \bibnamefont
			{Mikhailov}},\ }\href@noop {} {\bibfield  {journal} {\bibinfo  {journal}
			{Physical Review E}\ }\textbf {\bibinfo {volume} {79}},\ \bibinfo {pages}
		{036214} (\bibinfo {year} {2009})}\BibitemShut {NoStop}%
	\bibitem [{\citenamefont {Colizza}\ \emph {et~al.}(2007)\citenamefont
		{Colizza}, \citenamefont {Pastor-Satorras},\ and\ \citenamefont
		{Vespignani}}]{colizza2007reaction}%
	\BibitemOpen
	\bibfield  {author} {\bibinfo {author} {\bibfnamefont {V.}~\bibnamefont
			{Colizza}}, \bibinfo {author} {\bibfnamefont {R.}~\bibnamefont
			{Pastor-Satorras}},\ and\ \bibinfo {author} {\bibfnamefont {A.}~\bibnamefont
			{Vespignani}},\ }\href@noop {} {\bibfield  {journal} {\bibinfo  {journal}
			{Nature Physics}\ }\textbf {\bibinfo {volume} {3}},\ \bibinfo {pages} {276}
		(\bibinfo {year} {2007})}\BibitemShut {NoStop}%
	\bibitem [{\citenamefont {Colizza}\ and\ \citenamefont
		{Vespignani}(2008)}]{colizza2008epidemic}%
	\BibitemOpen
	\bibfield  {author} {\bibinfo {author} {\bibfnamefont {V.}~\bibnamefont
			{Colizza}}\ and\ \bibinfo {author} {\bibfnamefont {A.}~\bibnamefont
			{Vespignani}},\ }\href@noop {} {\bibfield  {journal} {\bibinfo  {journal}
			{Journal of Theoretical Biology}\ }\textbf {\bibinfo {volume} {251}},\
		\bibinfo {pages} {450} (\bibinfo {year} {2008})}\BibitemShut {NoStop}%
	\bibitem [{web(2025)}]{web}%
	\BibitemOpen
	\href@noop {} {}\bibinfo {howpublished} {\url
		{https://github.com/me-souravpamu/diffusion_ecoevo.git}} (\bibinfo {year}
	{2025})\BibitemShut {NoStop}%
	\bibitem [{\citenamefont {Hauert}\ \emph {et~al.}(2002)\citenamefont {Hauert},
		\citenamefont {De~Monte}, \citenamefont {Hofbauer},\ and\ \citenamefont
		{Sigmund}}]{hauert2002replicator}%
	\BibitemOpen
	\bibfield  {author} {\bibinfo {author} {\bibfnamefont {C.}~\bibnamefont
			{Hauert}}, \bibinfo {author} {\bibfnamefont {S.}~\bibnamefont {De~Monte}},
		\bibinfo {author} {\bibfnamefont {J.}~\bibnamefont {Hofbauer}},\ and\
		\bibinfo {author} {\bibfnamefont {K.}~\bibnamefont {Sigmund}},\ }\href@noop
	{} {\bibfield  {journal} {\bibinfo  {journal} {Journal of Theoretical
				Biology}\ }\textbf {\bibinfo {volume} {218}},\ \bibinfo {pages} {187}
		(\bibinfo {year} {2002})}\BibitemShut {NoStop}%
	\bibitem [{\citenamefont {Erdos}(1959)}]{erdos1959random}%
	\BibitemOpen
	\bibfield  {author} {\bibinfo {author} {\bibfnamefont {P.}~\bibnamefont
			{Erdos}},\ }\href@noop {} {\bibfield  {journal} {\bibinfo  {journal}
			{Mathematicae}\ }\textbf {\bibinfo {volume} {6}},\ \bibinfo {pages} {290}
		(\bibinfo {year} {1959})}\BibitemShut {NoStop}%
\end{thebibliography}

%

\onecolumngrid  
\appendix
\section*{Supplementary Information}

\setcounter{equation}{0}
\renewcommand{\theequation}{SI.\arabic{equation}}
\setcounter{figure}{0}
\renewcommand{\thefigure}{S\arabic{figure}}
\setcounter{table}{0}
\renewcommand{\thetable}{S\arabic{table}}

\section{Detailed construction of the eco-evolutionary public goods model (single population)}
\label{Model_construct}

In this section we derive the eco-evolutionary public goods model used in the main text. The derivation follows the ecological public-goods framework, where the population density determines the probability of forming interaction groups \cite{wakano2009spatial,wakano2011pattern,hauert2006evolutionary,hauert2008ecological}.

\subsection*{Population composition}

We consider a population composed of cooperators and defectors with densities $u$ and $v$, respectively. The remaining fraction corresponds to ecological vacancies,
$
z=1-u-v.
$
Vacant space represents empty sites where no individuals are present. Since both reproduction and interaction require occupied sites, the ecological vacancy directly affects the eco-evolutionary dynamics.

\subsection*{Group formation}

Public-goods interactions occur in groups of maximum size $N$. A focal individual is first selected, and the remaining $N-1$ positions are filled according to the population composition. Each position is occupied with probability $(1-z)$ and remains empty with probability $z$.

Suppose that the final interacting group contains $S$ individuals including the focal player. Then, among the remaining $N-1$ available positions, exactly $S-1$ positions must be occupied while the remaining $N-S$ positions remain empty. The number of possible ways to choose these $S-1$ occupied positions from the $N-1$ available positions is
$
\binom{N-1}{S-1}.
$
Therefore, the probability that the focal individual participates in a group of size $S$ is
\begin{equation}
\label{eq:NS}
P(S)=
\binom{N-1}{S-1}
(1-z)^{S-1}
z^{N-S}.
\end{equation}

Here, the factor $(1-z)^{S-1}$ gives the probability that $S-1$ positions are occupied, while $z^{N-S}$ gives the probability that the remaining $N-S$ positions are empty.

\subsection*{Payoff of defectors}

Consider now a focal defector participating in a group of size $S$. Among the remaining $S-1$ individuals, suppose that exactly $m$ are cooperators and the remaining $S-1-m$ are defectors.

The probability of observing exactly $m$ cooperators among the $S-1$ individuals follows a binomial distribution,
\[
\binom{S-1}{m}
\left(\frac{u}{u+v}\right)^m
\left(\frac{v}{u+v}\right)^{S-1-m}.
\]

The binomial coefficient $\binom{S-1}{m}$ counts the number of ways of selecting $m$ cooperators among the $S-1$ individuals. The probability that an occupied site contains a cooperator equals $u/(u+v)$, while the probability of finding a defector equals $v/(u+v)$.

Each cooperator contributes one unit to the public good. Therefore, if there are $m$ cooperators in the group, the total contribution equals $m$. After multiplication by the synergy factor $r$, the total public good becomes $rm$. Since the benefit is equally shared among all $S$ participants, the payoff obtained by the focal defector is
$
rm/S.
$

Averaging over all possible values of $m$ yields the expected payoff of defectors in a group of size $S$:
\begin{equation}
\label{eq:PD1}
P_D(S)=
\frac{r}{S}
\sum_{m=0}^{S-1}
m
\binom{S-1}{m}
\left(\frac{u}{u+v}\right)^m
\left(\frac{v}{u+v}\right)^{S-1-m}.
\end{equation}

To simplify this expression, we use the standard binomial identity
\[
\sum_{m=0}^{S-1}
m
\binom{S-1}{m}
x^m y^{S-1-m}
=
(S-1)x(x+y)^{S-2}.
\]

Setting
$
x=u/(u+v)
$
and
$
y=v/(u+v),
$
together with
$
x+y=1,
$
Eq.~\eqref{eq:PD1} reduces to
\[
P_D(S)
=
\frac{r}{S}(S-1)\frac{u}{u+v}.
\]

Therefore,
\begin{equation}
P_D(S)=
\frac{r(S-1)}{S}\frac{u}{u+v}.
\end{equation}

{\color{black}\subsection*{Payoff of cooperators}

Now consider a focal cooperator. If among the remaining $S-1$ individuals there are $m$ cooperators, then the total number of cooperators in the group equals $m+1$ because the focal individual also contributes.

The total contribution to the public good is therefore $m+1$, and after multiplication by the synergy factor $r$ the total benefit becomes $r(m+1)$. Since this amount is equally divided among all $S$ individuals, the focal cooperator receives
$
r(m+1)/S.
$

Because cooperators pay the cost of cooperation $c=1$, the payoff becomes
\[
\frac{r(m+1)}{S}-1.
\]

Separating the focal contribution from the defector payoff gives
\begin{equation}
P_C(S)=P_D(S)+\frac{r}{S}-1.
\end{equation}

\subsection*{Expected fitness}

Since interaction groups may have different sizes, the payoff of each strategy must be averaged over all possible group sizes. If $P_i(S)$ denotes the payoff of strategy $i\in\{C,D\}$ in a group of size $S$, then the expected fitness becomes
\[
f_i=
\sum_{S=2}^{N}
P(S)\,P_i(S),
\]
where $P(S)$ is given by ~\eqref{eq:NS}.

Substituting the explicit expression for $P_D(S)$ gives
\[
f_D=
\sum_{S=2}^{N}
\binom{N-1}{S-1}
(1-z)^{S-1}
z^{N-S}
\frac{r(S-1)}{S}\frac{u}{u+v}.
\]

Using
$
u+v=1-z,
$
we obtain
\[
f_D=
ru
\sum_{S=2}^{N}
\binom{N-1}{S-1}
\frac{S-1}{S}
(1-z)^{S-2}
z^{N-S}.
\]

Performing the summation yields
\begin{equation}
f_D=
\frac{ru}{1-z}
\left(
1-\frac{1-z^N}{N(1-z)}
\right).
\end{equation}

Similarly, the fitness of cooperators is obtained by averaging
$P_C(S)$ over all possible group sizes:
\[
f_C=
\sum_{S=2}^{N}
P(S)\,P_C(S).
\]

Substituting
$
P_C(S)=P_D(S)+\frac{r}{S}-1
$
gives
\[
f_C=
\sum_{S=2}^{N}
P(S)\left(P_D(S)+\frac{r}{S}-1\right).
\]

Using the definition of $f_D$, this becomes
\[
f_C=
f_D+
\sum_{S=2}^{N}
P(S)\left(\frac{r}{S}-1\right).
\]

Hence,
\begin{equation}
f_C=f_D-F(z),
\end{equation}
where
\[
F(z)=
\sum_{S=2}^{N}
P(S)\left(1-\frac{r}{S}\right).
\]

Substituting the explicit expression for the group-size probability,
\[
P(S)=
\binom{N-1}{S-1}
(1-z)^{S-1}
z^{N-S},
\]
we obtain
\[
F(z)=
\sum_{S=2}^{N}
\binom{N-1}{S-1}
(1-z)^{S-1}
z^{N-S}
\left(1-\frac{r}{S}\right).
\]

Separating the summation into two parts gives
\[
F(z)=
\sum_{S=2}^{N}
\binom{N-1}{S-1}
(1-z)^{S-1}
z^{N-S}
-r
\sum_{S=2}^{N}
\binom{N-1}{S-1}
\frac{(1-z)^{S-1}}{S}
z^{N-S}.
\]

Using the binomial identity
\[
\sum_{S=1}^{N}
\binom{N-1}{S-1}
(1-z)^{S-1}
z^{N-S}=1,
\]
the first summation becomes
\[
1-z^{N-1}.
\]

For the second summation, using
\[
\frac{1}{S}\binom{N-1}{S-1}
=
\frac{1}{N}\binom{N}{S},
\]
we obtain
\[
\sum_{S=2}^{N}
\binom{N-1}{S-1}
\frac{(1-z)^{S-1}}{S}
z^{N-S}
=
\frac{1}{N(1-z)}
\sum_{S=2}^{N}
\binom{N}{S}
(1-z)^S z^{N-S}.
\]

Applying the binomial expansion
\[
\sum_{S=0}^{N}
\binom{N}{S}
(1-z)^S z^{N-S}=1,
\]
and subtracting the $S=0$ and $S=1$ terms yields
\[
\sum_{S=2}^{N}
\binom{N}{S}
(1-z)^S z^{N-S}
=
1-z^N-N(1-z)z^{N-1}.
\]

Substituting this expression back gives
\[
F(z)=
1-z^{N-1}
-
\frac{r}{N(1-z)}
\left[
1-z^N-N(1-z)z^{N-1}
\right].
\]

Finally, simplifying the terms yields
\begin{equation}
F(z)=
1+(r-1)z^{N-1}
-\frac{r}{N}\frac{1-z^N}{1-z}.
\end{equation}

The function $F(z)$ represents the effective reduction in cooperative payoff caused by ecological vacancies and incomplete interactions.}

\subsection*{Eco-evolutionary dynamics}

Population densities evolve according to their reproductive success. Reproduction requires available ecological space and therefore occurs at rate
$
z(f_{C,D}+b),
$
where $b$ is a baseline birth rate ensuring population viability even when payoff contributions are weak.

Let $d$ denote the constant death rate. Combining ecological reproduction and mortality yields the eco-evolutionary dynamics
\begin{equation}
\label{eq:econospace}
\begin{aligned}
\dot{u} &= u\left(z(f_C+b)-d\right),\\
\dot{v} &= v\left(z(f_D+b)-d\right).
\end{aligned}
\end{equation}

These equations describe the coupled ecological and evolutionary dynamics of cooperators and defectors in a single well-mixed population and form the basis for the spatial and network extensions used in the main text.

\section{Stability analysis of the model} \label{stability_analysis}

Before introducing diffusion across the network, it is instructive to analyze the local eco--evolutionary dynamics of a single patch. In the absence of spatial coupling, the model reduces to the ordinary differential equations~\eqref{eq:econospace}. The latter system corresponds to Eq.~(2) in the main text and provides the reference state for the analysis of the spatially extended model, Eq.~(3).

\subsection*{Equilibria}

The system admits two equilibria of particular interest.

\paragraph{Extinction equilibrium.}
The trivial equilibrium
\[
O=(0,0)
\]
represents the extinction of the population. Linear stability analysis shows that $O$ is stable whenever $b<d$.

\paragraph{Coexistence equilibrium.}
A second equilibrium corresponds to coexistence between cooperators and defectors,

\[
Q=(u^*,v^*), \qquad u^*>0,\; v^*>0 .
\]

This equilibrium is obtained by solving

\begin{equation}
z^*(f_C^*+b)-d=0, \qquad z^*(f_D^*+b)-d=0 .
\end{equation}

Since by definition $f_C=f_D-F(z)$, these relations imply

\begin{equation}
F(z^*)=0 .
\end{equation}

Hence the existence of the coexistence equilibrium is given by the roots of $F(z)$. The existence of $Q$ requires the condition $r>2$, because otherwise $F(z)>0$ and no interior solution exists. Whenever it exists, the equilibrium $Q$ is unique in $[0,1)$.

{\color{black} \subsection*{Existence and uniqueness of the coexistence equilibrium}

We now briefly discuss the condition under which the equation
\[
F(z)=0
\]
admits a biologically relevant solution in the interval $z\in(0,1)$.
Following Ref.~\cite{hauert2002replicator}, it can be shown that no such solution exists for $r\le2$, whereas for $r>2$ a unique root exists in $(0,1)$.

To examine the behavior of the function near the boundary of the admissible interval, it is convenient to define
\[
G(z)=F(z)(1-z).
\]
Since $1-z>0$ for $z\in(0,1)$, the functions $F(z)$ and $G(z)$ possess the same roots within this interval. Substituting the expression for $F(z)$ yields
\begin{equation}
\label{eq:defGz}
G(z)
=
(1-z)\left[1+(r-1)z^{N-1}\right]
-\frac{r}{N}(1-z^N).
\end{equation}

Evaluating at $z=0$ gives
\[
G(0)=1-\frac{r}{N}.
\]
Because the public-goods interaction requires $r<N$, we have
\[
G(0)>0.
\]
Consider next the limit $z\to1$. Writing
\[
z=1-\varepsilon,
\qquad \varepsilon\to0^+,
\]
and expanding in powers of $\varepsilon$ gives (once rewritten in the variable $z$)
\[
G(z)
\sim
\frac{(2-r)(N-1)}{2}(1-z)^2.
\]
Hence,
\begin{equation*}
G(z)>0
\quad \text{near } z=1
\quad \text{if } r<2,
\end{equation*}
whereas
\[
G(z)<0
\quad \text{near } z=1
\quad \text{if } r>2 \, .
\]


Therefore, the sign change near $z=1$ indicates that coexistence solutions may arise only for sufficiently large multiplication factors, i.e., $r>2$. A rigorous proof that the equation $F(z)=0$ possesses no roots for $r\le2$ and exactly one root for $r>2$ is given in Ref.~\cite{hauert2002replicator}.}

\begin{figure*}[hpt]
		\centering
\includegraphics[width=\linewidth]{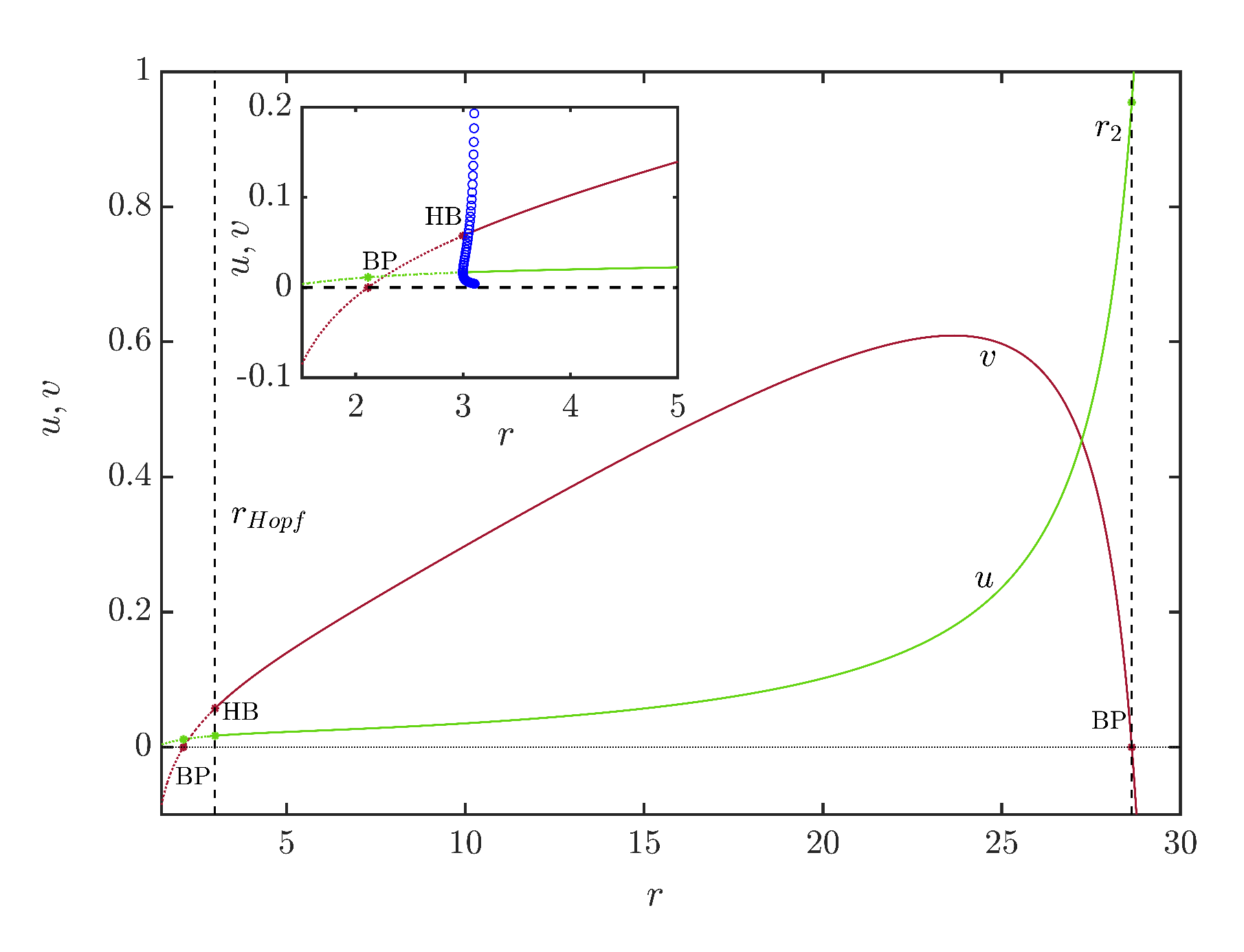} 
	\caption{\textbf{Bifurcation structure of the diffusion-free eco-evolutionary public-goods system.} Numerical continuation of the diffusion-free system with respect to the multiplication factor $r$, obtained using MATCONT for the parameter set $N=30$, $b=1$, and $d=1.3$. The green and red curves represent the equilibrium densities of cooperators $(u)$ and defectors $(v)$, respectively. Solid branches denote stable equilibria, whereas dashed branches correspond to unstable equilibria. The vertical dashed lines indicate the critical thresholds $r_{\mathrm{Hopf}}=2.9872$ and $r_2=28.6382$. A branch point bifurcation (BP) gives rise to the coexistence equilibrium at low values of $r$, while a Hopf bifurcation (HB) destabilizes the coexistence state near $r=r_{\mathrm{Hopf}}$. The inset shows the local bifurcation structure near the onset of coexistence and the Hopf transition. Blue circles correspond to unstable periodic orbits emerging from the Hopf point, consistent with a subcritical Hopf bifurcation. For $r_{\mathrm{Hopf}}<r<r_2$, both species coexist in a locally stable state. As $r$ increases further, the cooperator density increases monotonically whereas the defector density eventually decreases and vanishes at $r=r_2$, beyond which the system approaches a purely cooperative state. Throughout the continuation analysis, all biologically relevant solutions remain confined within the admissible simplex $u+v\le1$.
}
\label{bifurcation}
\end{figure*}

{\color{black} \section*{Stability properties and dynamical regimes of the diffusion-free system}

In this section we analyze the local stability properties of the diffusion-free eco-evolutionary system for the parameter regime used throughout the main text, namely $r=4$, $N=30$, $b=1$, and $d=1.3$. The qualitative behavior discussed below, including the existence of a stable coexistence equilibrium and the emergence of a Hopf bifurcation as $r$ varies, depends on the choice of parameter values and should not be interpreted as a universal property of the model. The stability of $Q$ depends strongly on the multiplication factor $r$ of the public good and on the demographic parameters $b$ and $d$. Further, we choose the regime $b\ge1$ and $d<b$, which ensures that defectors cannot survive in the absence of cooperators ($\dot v<0$ for $u=0$). Similar parameter choices have been used in previous studies of eco-evolutionary public-goods dynamics to investigate diffusion-driven pattern formation.

The existence and stability of the equilibrium states are summarized in Fig.~\ref{bifurcation}, obtained using numerical continuation with MATCONT by treating the multiplication factor $r$ as the bifurcation parameter. The continuation analysis confirms the existence of two critical thresholds, namely the Hopf bifurcation point $r_{\mathrm{Hopf}}=2.9872$ and the second threshold $r_2=28.6382$, indicated by the vertical dashed lines in Fig.~\ref{bifurcation}. The green and red branches represent the equilibrium densities of cooperators $(u)$ and defectors $(v)$, respectively. Solid curves denote stable equilibrium branches, whereas dotted curves correspond to unstable branches. The inset highlights the local bifurcation structure near the onset of coexistence and the Hopf transition.

For small multiplication factors, the extinction equilibrium remains stable and no biologically relevant coexistence state exists. As $r$ increases, a coexistence branch emerges through a branch point bifurcation (BP) near $r\approx2.11$, after which both cooperative and defector densities become positive. Initially this coexistence branch is unstable, as indicated by the dotted curves in the inset of Fig.~\ref{bifurcation}. At the critical value $r=r_{\mathrm{Hopf}}=2.9872$, the coexistence equilibrium undergoes a Hopf bifurcation (HB). The blue circles in Fig.~\ref{bifurcation} shown in the inset correspond to unstable periodic orbits generated near the Hopf point, consistent with a subcritical Hopf bifurcation. Consequently, stable periodic oscillations do not emerge in the parameter regime considered here.

For the interval $r_{\mathrm{Hopf}}<r<r_2$, the coexistence equilibrium becomes locally stable and both species persist. In this regime, the cooperator density increases monotonically with increasing $r$, reflecting the enhanced benefit generated by cooperative interactions. The defector density also initially increases with $r$, reaches a maximum at intermediate values of the multiplication factor, and subsequently decreases as cooperation becomes increasingly advantageous. Although the coexistence equilibrium is locally stable in this regime, its basin of attraction remains limited. Initial conditions with sufficiently low total population density or with an excessively large fraction of defectors may still drive the system toward extinction, indicating an Allee-type effect in the eco-evolutionary dynamics.

At the second threshold $r_2=28.6382$, the defector branch terminates through another branch point bifurcation. Beyond this threshold the system approaches a purely cooperative state characterized by $v\rightarrow0$ and $u\rightarrow1$. Importantly, throughout the continuation analysis all biologically relevant solutions remain confined within the admissible simplex $u+v\le1$, ensuring that the equilibrium densities always remain within the physically meaningful ecological domain of the model.}






\subsection*{Phase portrait}

\begin{figure*}[h]
		\centering
\includegraphics[width=0.8\linewidth]{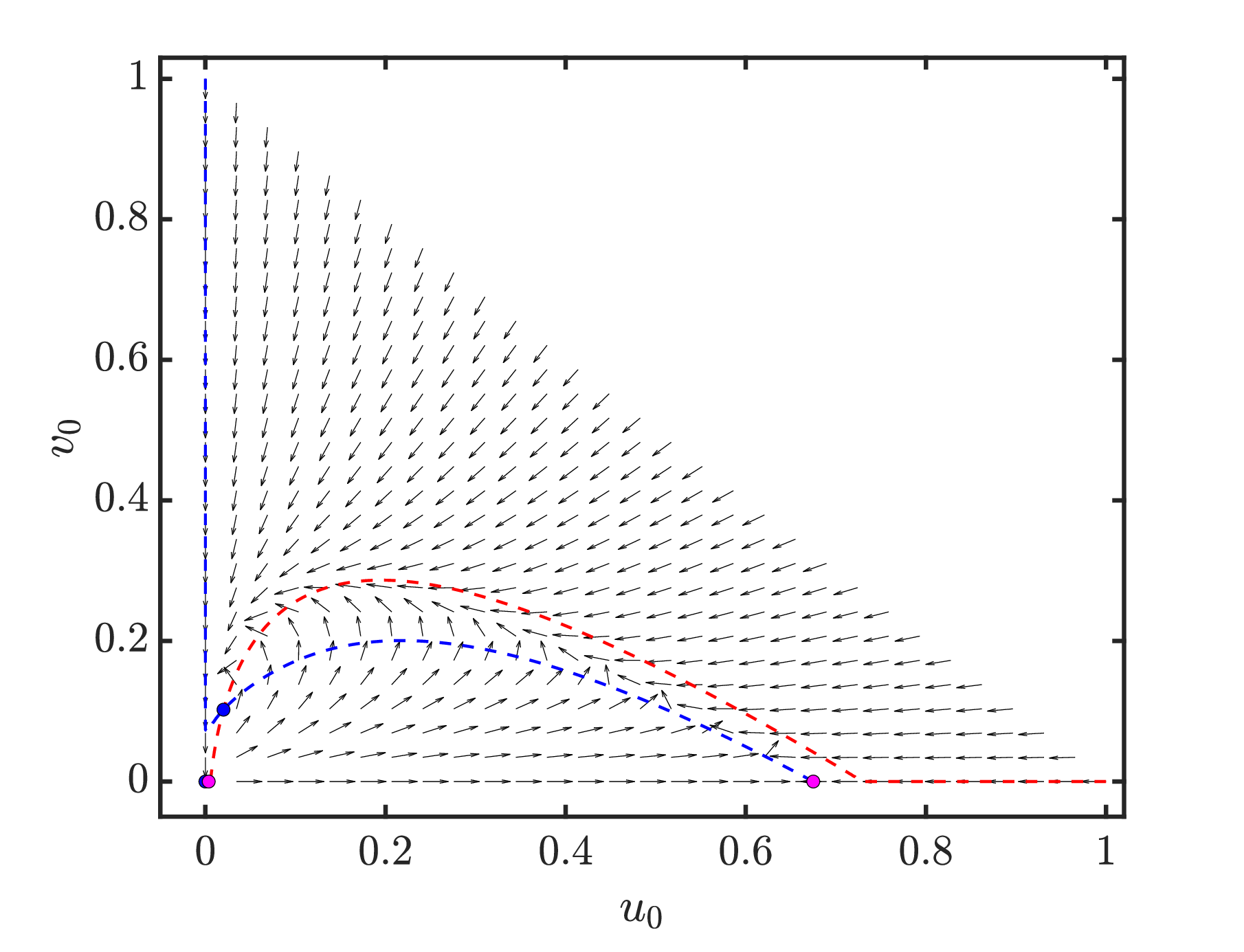} 
	\caption{\color{black}
\textbf{Phase Portrait and Stability Structure of the Diffusion-Free System:}
Phase portrait of the diffusion-free system on the admissible simplex $u+v \leq 1$ for $r=4$, $N=30$, $b=1$, and $d=1.3$. Black arrows indicate the direction of the phase flow $(\dot{u},\dot{v})$, with vectors normalized to emphasize direction rather than magnitude. The blue dashed curve corresponds to the nullcline $\dot{u}=0$, while the red dashed curve represents the nullcline $\dot{v}=0$. Their intersections determine the equilibrium points. Blue filled circles indicate stable equilibria, whereas pink filled circles denote saddle equilibria. The interior coexistence equilibrium $Q=(0.02,0.102)$ is a spirally stable fixed point. Along the $u$-axis, $(0,0)$ is stable, while $(0.00368,0)$ and $(0.675,0)$ are saddle points. The stable and unstable manifolds of the saddle equilibria generate the separatrix structure that partitions the basins of attraction within the simplex.
}
\label{equilibria}
\end{figure*}

In the parameter regime used in our study (e.g., $r = 4$, $N = 30$, $b = 1$, and $d = 1.3$)~\cite{wakano2009spatial,hauert2002replicator}, the diffusion-free system admits several equilibrium states, whose locations, eigenvalues, and stability properties are summarized in Table~\ref{tab:fixedpoints}. Among these, only the equilibria satisfying $u\ge0$, $v\ge0$, and $u+v\le1$ are biologically admissible and therefore dynamically relevant. The corresponding phase portrait within the admissible simplex $u+v\le1$ is shown in Fig.~\ref{equilibria}. The black arrows represent the normalized phase flow $(\dot u,\dot v)$ and therefore indicate only the direction of trajectories for different initial conditions.

The dashed curves denote the nullclines of the system: the blue dashed curve corresponds to $\dot u = 0$, while the red dashed curve corresponds to $\dot v = 0$. Their intersections determine the equilibrium points of the system. Stability is encoded directly in Fig.~\ref{equilibria}: blue filled circles indicate stable equilibria (all eigenvalues having negative real parts), whereas pink filled circles denote saddle equilibria.

The interior equilibrium 
\(
Q=(0.020078,\,0.102254)
\)
appears as a stable spiral coexistence state with eigenvalues
\(
-0.0707477 \pm 0.460856\, i,
\)
indicating damped oscillatory relaxation toward coexistence. This equilibrium attracts most interior initial conditions within the simplex. Along the $u$-axis, three biologically admissible boundary equilibria are also present. The trivial extinction equilibrium $(0,0)$ is a stable node with eigenvalues $\{-0.3,-0.3\}$ since $b<d$. In contrast, the equilibria $(0.00368781,0)$ and $(0.675,0)$ are saddle points. Due to the scale of Fig.~\ref{equilibria}, the points $(0,0)$ and $(0.00368781,0)$ lie extremely close to one another and are therefore nearly indistinguishable visually.

Table~\ref{tab:fixedpoints} additionally contains two mathematically obtained equilibria, namely $(-1.1672,3.12842)$ and $(0,-0.3)$. Since these solutions involve negative population densities, they lie outside the admissible ecological domain and are therefore biologically irrelevant. Consequently, they are omitted from the phase portrait shown in Fig.~\ref{equilibria}.

The phase portrait further reveals the separatrix structure generated by the saddle manifolds, which partitions the basins of attraction. The basin of attraction of the extinction equilibrium $(0,0)$ remains comparatively small and confined near the boundary of the simplex, whereas the stable coexistence equilibrium $Q$ attracts most interior trajectories. Hence, under diffusion-free dynamics, the system predominantly converges toward stable coexistence, providing the homogeneous reference state from which diffusion-driven heterogeneous patterns emerge in the networked model.

{\color{black}
\begin{table}[h!]
\centering
\caption{\color{black}{Equilibrium points of the diffusion-free system for $r=4$, $N=30$, $b=1$, and $d=1.3$, together with the corresponding eigenvalues and local stability properties. Only equilibria satisfying $u\ge0$, $v\ge0$, and $u+v\le1$ are biologically admissible.}}

\label{tab:fixedpoints}
\begin{tabular}{c c c c}
\hline
$u^{*}$ & $v^{*}$ & Eigenvalues & Stability \\
\hline
$0$ & $0$ & $\{-0.3,\;-0.3\}$ & stable \\

$-1.1672$ & $3.12842$ & $\{28.1122,\;6.79028\}$ & Unstable \\

$0.675$ & $0$ & $\{-2.7,\;0.260802\}$ & Saddle \\

$0$ & $-0.3$ & $\{-6348.09,\;0.3\}$ & Saddle \\

$0.020078$ & $0.102254$ & $\{-0.0707477 \pm 0.460856 i\}$ & Stable \\

$0.00368781$ & $0$ & $\{0.283429,\;-0.0977392\}$ & Saddle \\
\hline
\end{tabular}
\end{table}}

{\color{black}
\section{Perspectives of breaking the co-existence phenomenon of strategic individuals in different network topologies}
In the main text of our study, we have already observed and analysed the emergence of quantitively different behavior, resulting into pattern phenomenon from the steady co-existence of the competitive individuals performing in the N-player public goods dilemma, embedded in the BA scale-free network of 1000 nodes, and $\langle k\rangle=10$. Here, we observe and analyse the pattern formation phenomenon of the strategic individuals being connected by other network topologies (e.g. the Erd\H{o}s-R\'enyi random network and the Watts-Strogatz small world network, respectively). All dynamical parameters are identical to those used
in the main text ($r=4$, $N=30$, $b=1$, $d=1.3$, $\epsilon=0.01$,
and $\kappa=100$).

 \begin{figure*}[h]
		\centering
        \includegraphics[width=0.8\linewidth]{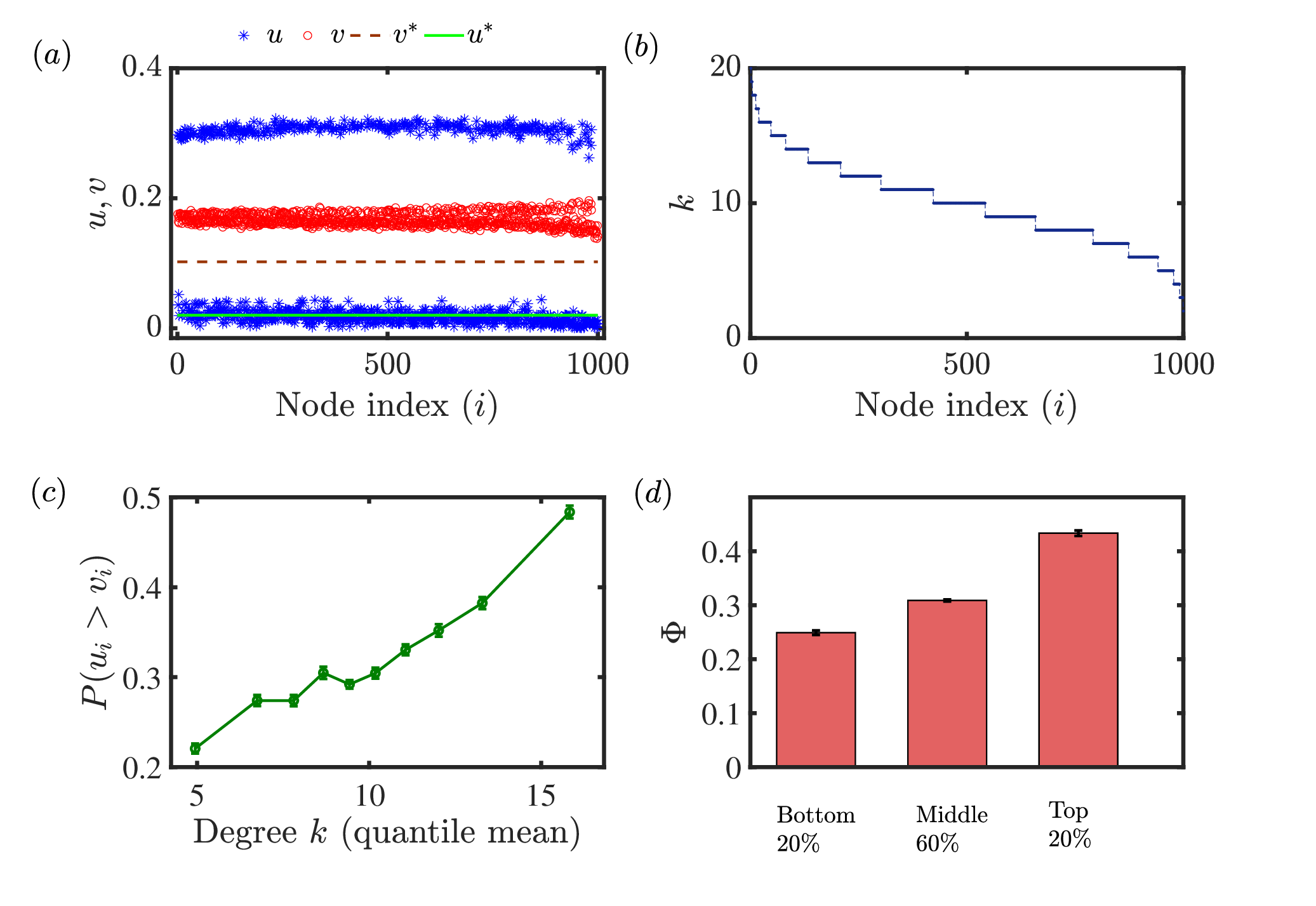} 
	\caption{ \color{black}
\textbf{Diffusion-induced heterogeneous cooperation patterns in an Erd\H{o}s–Rényi random network.}
($a$) Stationary densities of cooperators $u_i$ (blue) and defectors $v_i$ (red) across nodes ordered by decreasing degree. Horizontal lines indicate the homogeneous equilibrium $(u^*,v^*)$ of the aspatial system. With asymmetric mobility $(\epsilon=0.01,\kappa=100)$ diffusion breaks the uniform state and generates heterogeneous densities across nodes. ($b$) Degree sequence of the Erd\H{o}s–Rényi network ($M=1000$, $\langle k\rangle=10$). In contrast to the heavy-tailed scale-free topology, the degrees are narrowly distributed around the mean, indicating the absence of hubs. ($c$) $P\{(u_i>v_i)\mid k\}$, probability that a node becomes cooperator-dominated as a function of node degree. Nodes are grouped using quantile-based binning; points represent averages over $100$ realizations with standard error bars. The probability increases with degree but more gradually than in the scale-free case. ($d$) Degree-stratified fraction $(\Phi)$ of cooperator-dominated nodes for the bottom $20\%$, middle $60\%$, and top $20\%$ of the degree distribution. Unlike the scale-free network, the top $20\%$ does not exhibit a sharply larger $\Phi$, reflecting the limited degree heterogeneity of random networks.}
		\label{random1}
     \end{figure*}
     
\subsection{Diffusion-driven instability on a random network}

To examine whether the diffusion-driven mechanism observed in the
scale-free topology persists in networks with weaker structural
heterogeneity, we repeat the numerical experiments on an Erd\H{o}s-R\'enyi random network \cite{erdos1959random}. The network consists of $M=1000$ nodes with average degree
$\langle k\rangle =10$, generated by connecting node pairs with equal
probability. In contrast to the scale-free topology analyzed in
Fig.~2 of the main text, random networks do not exhibit
strong degree heterogeneity and therefore lack highly connected nodes. Instead, most nodes possess similar degrees clustered around the network average.

The stationary node-level densities obtained after the diffusion-driven instability are shown in Fig.~\ref{random1}$(a)$. Blue stars and red circles represent the asymptotic densities of cooperators ($u_i$) and defectors ($v_i$), respectively, while the horizontal lines denote the homogeneous equilibrium values $(u^*,v^*)$ of the aspatial system. Consistently with the behavior observed in the scale-free case, asymmetric diffusion destabilizes the homogeneous coexistence and generates heterogeneous stationary densities across the network. Cooperators increase substantially above their aspatial equilibrium
value, and many nodes exhibit cooperative dominance ($u_i>v_i$) even
though the isolated dynamics favors defectors.

However, the spatial organization differs markedly from the scale-free case. The connectivity structure of the random network, shown in
Fig.~\ref{random1}$(b)$, reveals a narrow distribution of node degrees
with no extremely highly connected nodes. As a consequence, the
heterogeneous pattern does not display the strong structural
stratification observed in Fig.~2 of the main text. Instead, cooperative and defector-dominated nodes are distributed more
uniformly across the network.

Despite the absence of pronounced structural heterogeneity, the
dependence of cooperation on node connectivity remains detectable.
This relationship is quantified in Fig.~\ref{random1}$(c)$, where nodes are grouped using quantile-based binning of the degree distribution and the vertical axis represents the probability that a node is cooperator-dominated within each degree class. The probability $P(u_i>v_i)$ increases systematically with node degree, indicating that nodes with comparatively larger connectivity are more likely to sustain cooperative dominance.

A complementary comparison is provided in Fig.~\ref{random1}($d$), where nodes are divided into three percentile classes of the degree distribution: the bottom $20\%$, the middle $60\%$, and the top $20\%$. The fraction $\Phi$ of nodes satisfying $u_i>v_i$ again increases with degree class, confirming a positive association between node connectivity and cooperative dominance. However, in contrast to the scale-free network analyzed in the main text, the differences between the three degree classes are less pronounced. In scale-free networks, the broad degree distribution produces highly connected hubs, leading to a clear separation where the top $20\%$ of nodes exhibit substantially higher cooperative dominance. In the Erdős–Rényi network considered here, the degree distribution is much narrower, so the distinction between high- and low-degree nodes is weaker. Consequently, the fractions of cooperative nodes across the three groups differ only moderately. Nevertheless, because finite random networks still exhibit small variations in node degree, nodes with comparatively larger connectivity remain more likely to sustain cooperative dominance than those with lower degree.

These results highlight an important distinction between network topologies. In the scale-free network of the main text, the broad degree distribution produces pronounced structural differentiation, leading to strongly cooperative hubs. In contrast, the Erdős–Rényi network exhibits only moderate degree variability, resulting in a more homogeneous spatial organization. Nevertheless, even this limited variability is sufficient for the diffusion mechanism to bias cooperation toward nodes with relatively larger connectivity. This comparison indicates that while strong degree heterogeneity amplifies the emergence of cooperative hubs, the underlying diffusion-driven mechanism promoting cooperation remains operative across different network architectures.

\subsection{Diffusion-driven instability on a small-world network}

To examine further the robustness of the diffusion-driven mechanism, we repeat the numerical experiments on a Watts--Strogatz small-world network \cite{watts1998collective}. The network is generated with
$M=1000$ nodes, average degree $<k>=10$, and rewiring probability
$p_{sw}=0.15$. 

The stationary node-level densities obtained after the diffusion-driven instability are shown in Fig.~\ref{small-world1}. Blue stars and red circles denote the asymptotic densities of cooperators ($u_i$) and defectors ($v_i$), respectively. The horizontal lines indicate the homogeneous equilibrium values $(u^*,v^*)$ of the aspatial system. As in the scale-free and random network, diffusion destabilizes the homogeneous coexistence state and produces heterogeneous stationary densities across nodes. In particular, many nodes exhibit cooperative dominance ($u_i>v_i$), even though the isolated eco-evolutionary dynamics favors defectors.

The heterogeneous stationary distribution shown in Fig.~\ref{random1} reveals that, in contrast to the scale-free network, the degree-dependent stratification of cooperators is much less pronounced. Nodes exhibiting high cooperative densities are distributed more uniformly across the network rather than being concentrated among specific high-degree nodes. This behavior reflects the structural properties of the small-world topology, where the degree is narrowly centered around the average degree $<k>=10$ due to the construction of the Watts–Strogatz network, leading to weak heterogeneity in node connectivity. As a result, nodes experience comparable diffusive coupling strengths, and the mobility-induced spatial differentiation becomes more homogeneous across the network. Nevertheless, asymmetric mobility still enhances the overall abundance of cooperators relative to the aspatial equilibrium, demonstrating that diffusion can break homogeneous coexistence even in the absence of strong hubs. The extent of this effect, however, is reduced compared to scale-free networks, highlighting the role of connectivity heterogeneity in amplifying cooperative dominance. Since small-world networks exhibit only weak degree variability, the degree-resolved measures analogous to Fig.~\ref{random1}$(b)$--$(d)$ do not provide qualitatively new insights and closely resemble those obtained for Erdős–Rényi random networks; therefore, these panels are omitted for brevity.


 \begin{figure*}[htp]
		\centering
        \includegraphics[width=0.6\linewidth]{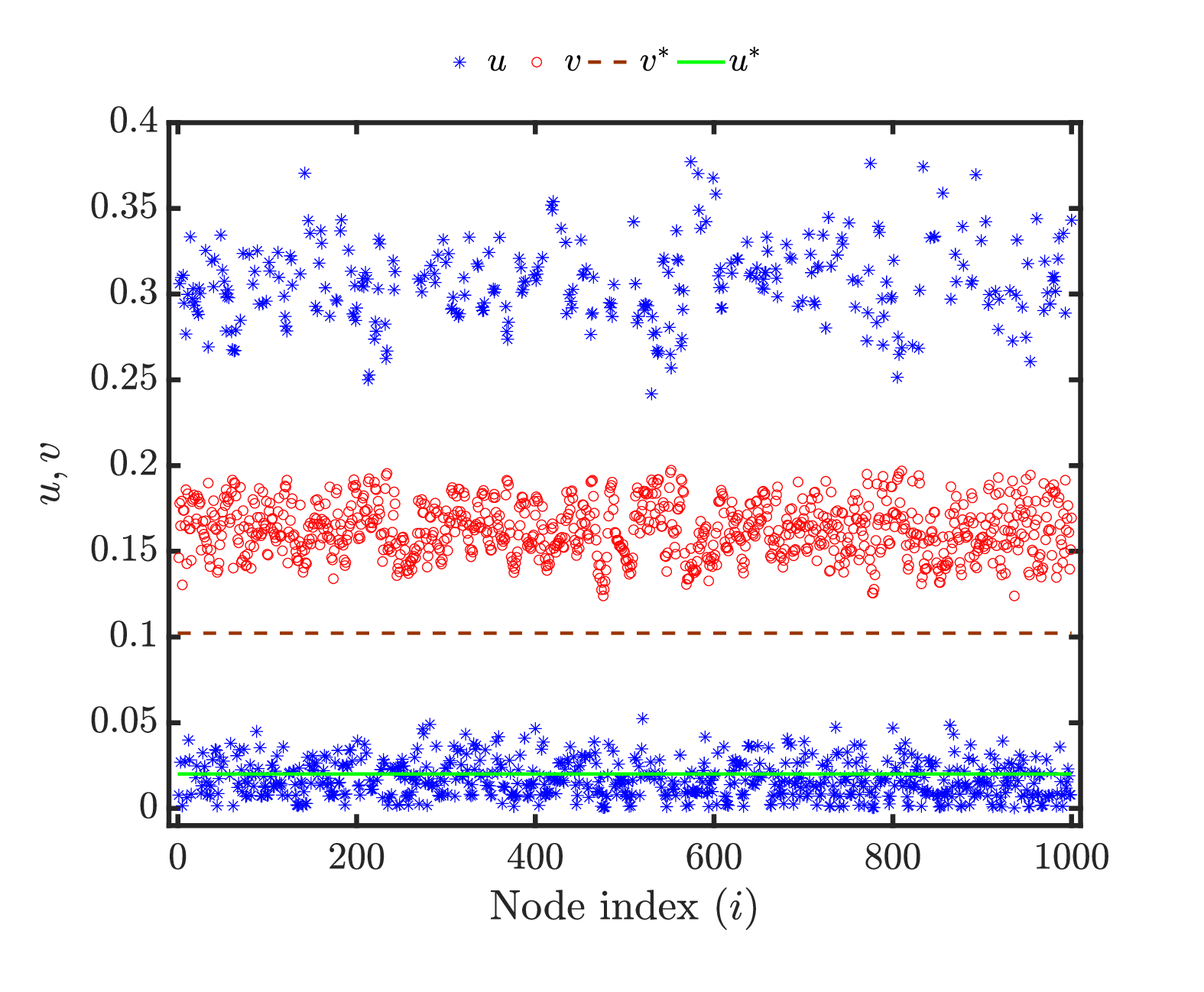} 
	\caption{\color{black}
\textbf{Diffusion-induced heterogeneous cooperation patterns in a small-world network.}
Stationary densities of cooperators $u_i$ (blue crosses and defectors $v_i$ (red circles) across nodes ordered by decreasing degree. Horizontal lines denote the homogeneous equilibrium $(u^*,v^*)$ of the aspatial system. With asymmetric mobility $(\epsilon=0.01,\kappa=100)$ diffusion breaks the uniform coexistence and generates heterogeneous densities across nodes.}
		\label{small-world1}
     \end{figure*}

}

{

\section{ Diffusion-induced destabilization in the $(r, N)$ plane: Linking group size to instability Patterns}

 \begin{figure*}[h]
		\centering
        \includegraphics[width=0.8\linewidth]{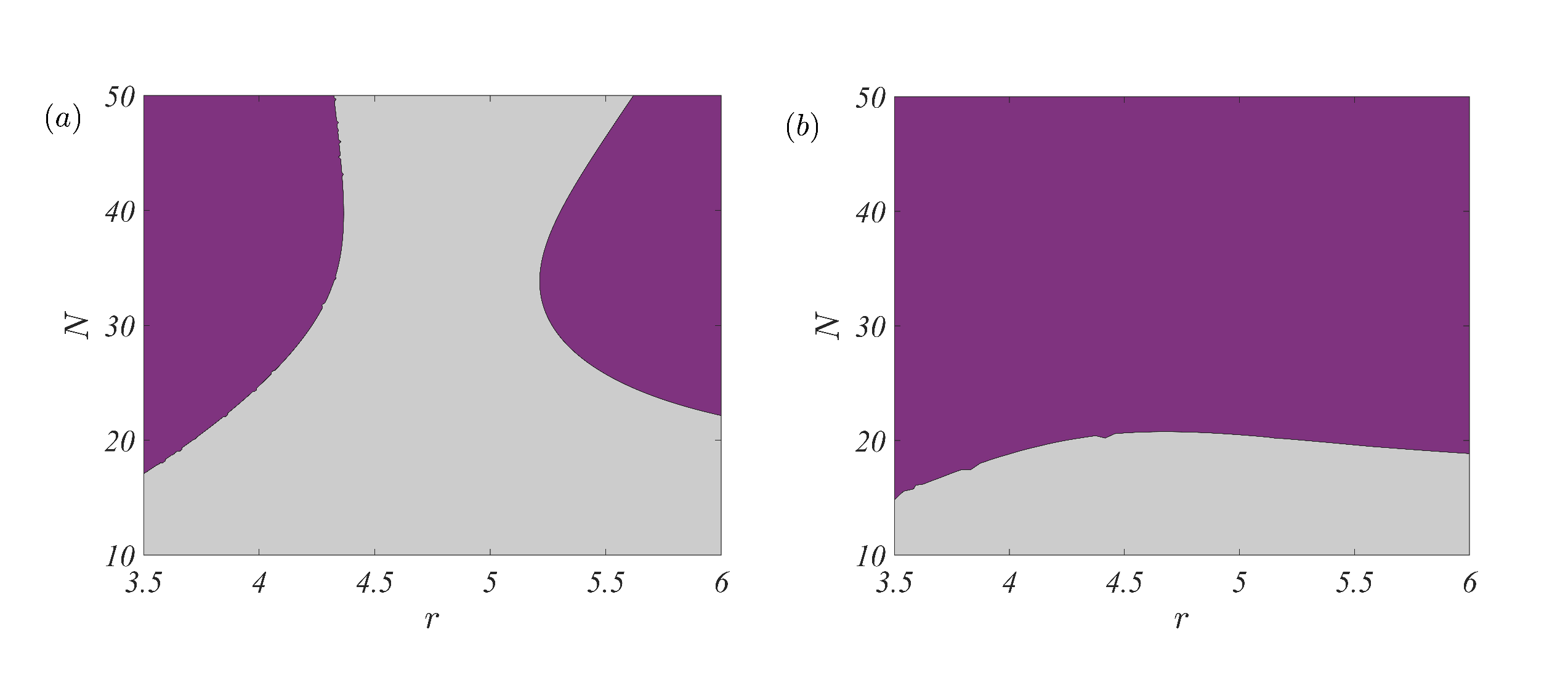} 
	\caption{
\textbf{Diffusion-induced instability in the $(r, N)$ plane.}
We report phase diagrams showing the maximum of the dispersion relation $\Lambda^{\max}$; we can observe an interesting separation of unstable cases $\Lambda_{\max} > 0$ (purple) and stable ones $\Lambda_{\max} < 0$ (grey), highlighting regions where spatial heterogeneity emerges due to diffusion-induced destabilization in the eco-evolutionary public goods game.The parameter values used for this experiment are $(a)$ $\kappa = 18$, and $(b)$  $\kappa = 20$ at $\epsilon = 0.01$.
}
		\label{r_n_analytics}
     \end{figure*}

To analytically capture the way in which the size of the interaction group $N$ interplays with the synergy factor $r$ to induce diffusion-driven instability, we examine the maximum of the dispersion relation $\Lambda^{\mathrm{max}}$ by focusing on the coupling between ecological population effects (via $N$) and evolutionary game dynamics (via $r$), under spatial diffusion. Fig. \ref{r_n_analytics} presents the $(r, N)$-phase diagrams of $\Lambda^{\mathrm{max}}$, by identifying the stability landscape in multiple regimes of sensitivity parameter $\kappa$ and a fixed diffusion strength $\epsilon$.

Figure \ref{r_n_analytics}$(a)$ and \ref{r_n_analytics}($b$) show the system with low mobility $\epsilon = 0.01$ and moderate nonlinear sensitivity $\kappa = 18$ and $\kappa = 20$, respectively. In this weak-diffusion regime, the system remains largely stable (gray) for small $N$, regardless of $r$, as the influence of any single individual remains localized. However, as $N$ increases, a non-monotonic destabilization pattern emerges. This behavior originates from the fact that, in the proposed eco-evolutionary public goods game formulation, increasing $N$ dilutes individual returns from cooperation ($\frac{r}{N}$) while simultaneously increasing the collective contribution pool. The Jacobian reaction terms scale non-linearly with $N$, and so does the diffusion-driven perturbations, leading to sharp transitions in the eigenvalue spectrum for intermediate group sizes, as testified by the boundaries of the domains.

These destabilizations are a direct manifestation of spatial symmetry breaking: when cooperative gains scale inadequately with $N$, while spatial heterogeneity allows local subgroups of defectors to outcompete their neighbors, the homogeneous state destabilizes. Moreover, due to the spatially non-uniform payoff gradients embedded in the diffusion term, eigenvalues can locally become positive even if the system appears stable under well-mixed dynamics. Notably, the difference between $\kappa = 18$ and $\kappa = 20$ in Fig. \ref{r_n_analytics}($a$) and \ref{r_n_analytics}($b$) highlights the sensitivity of the system stability to the rate at which players adjust their strategies based on local payoff differences. Even a small change in $\kappa$ shifts the unstable region along both $r$ and $N$, signaling a strong eco-evolutionary feedback loop driven by behavioral plasticity.

\color{black}
\section{Effect of the diffusion strength on the instability interval}
\label{app:interval}
\begin{figure*}[h]
		\centering
\includegraphics[width=0.5\linewidth]{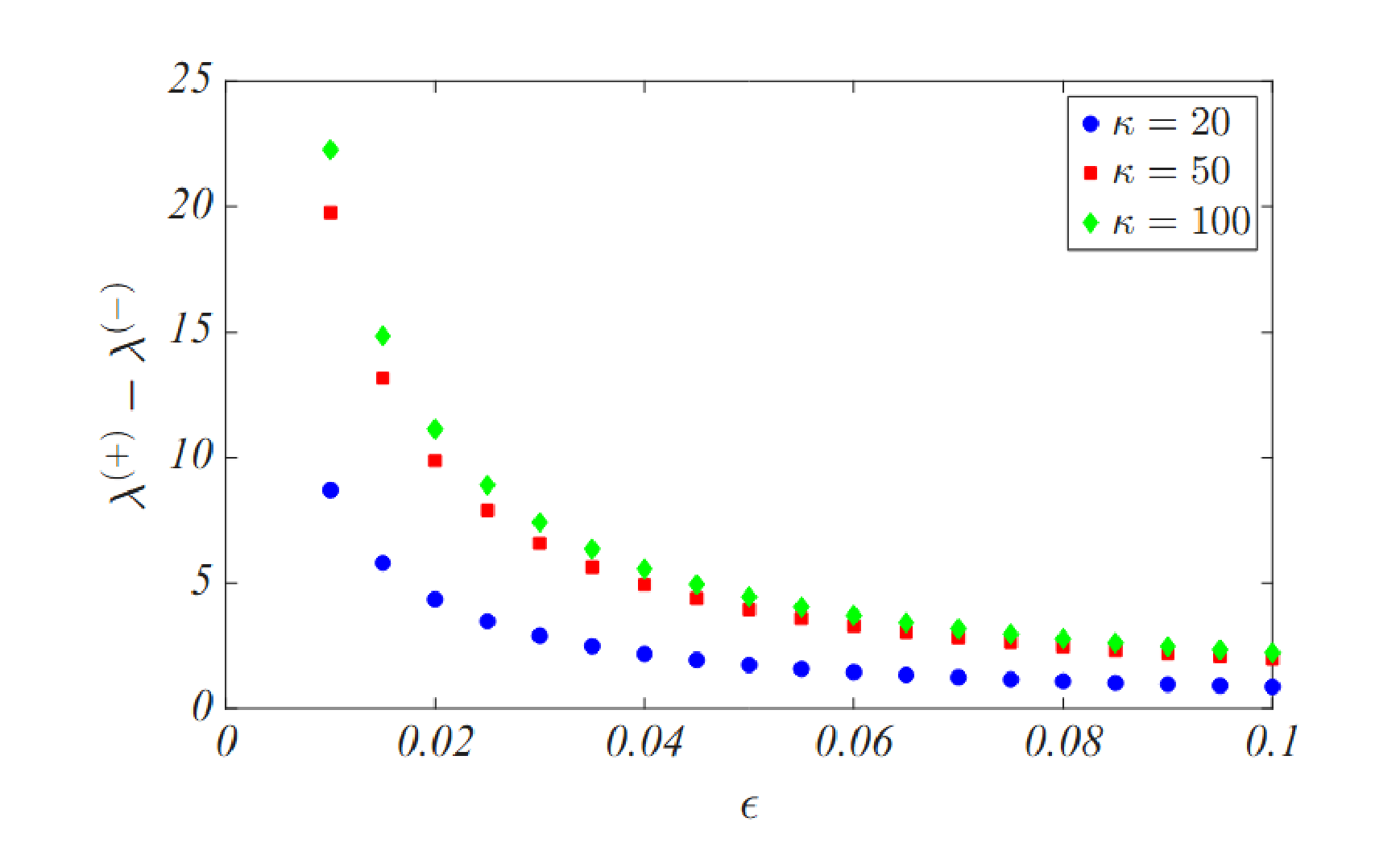} 
	\caption{
\textbf{The size of the instability interval}, $\lambda^{(+)}-\lambda^{(-)}$ as a function of $\epsilon$ for few values of $\kappa$ (blue circles $\kappa=20$, red squares $\kappa=50$ and green diamonds $\kappa=100$).
}
\label{fig:interval}
\end{figure*}
In the main text (Section 4.1 of Materials and Methods), we derived the analytical condition for the critical
mobility ratio $\kappa_c$ marking the onset of diffusion-driven
instability. Here we provide additional insight into how the diffusion
strength $\epsilon$ and the mobility ratio $\kappa$ influence the
range of Laplacian modes that can destabilize the homogeneous
equilibrium.

When $\kappa>\kappa_c$, the dispersion relation becomes positive only
within a finite interval of Laplacian eigenvalues,
\[
\lambda^{(-)}<\lambda_\alpha<\lambda^{(+)} .
\]
The width of this interval, $\lambda^{(+)}-\lambda^{(-)}$, determines
how many Laplacian modes may potentially trigger spatial
heterogeneity. The larger the interval, the higher the probability
that eigenvalues of the network Laplacian fall within the unstable
region.

Figure~\ref{fig:interval} illustrates the dependence of the interval
width on the diffusion strength $\epsilon$ for different values of the
mobility ratio $\kappa$. The blue circular markers correspond to
$\kappa=20$, the red square markers to $\kappa=50$, and the green
diamond markers to $\kappa=100$.

The results reported in Fig.~\ref{fig:interval} show that the smaller
$\epsilon$ the larger the interval, indeed the size of the latter
increases as $1/\epsilon$. At the same time, for a fixed $\epsilon$,
the larger $\kappa$, the larger the interval and for $\kappa \gg 1$ we
get $\lambda^{(+)}-\lambda^{(-)}\sim |f_u|/\epsilon$.

Finally, let us remark that the discrete nature of the Laplace spectrum,
induced by the finite size of the network, can create a novel phenomenon
that highlights the impact of finite-size and topological effects: while
the analytical study captures the universal instability threshold given,
i.e., by a positive $\lambda(x)$, the network structure can substantially
modify the emergence of patterns, by preventing the latter ones to
emerge, because of a negative dispersion relation once evaluated on the
discrete Laplace spectrum.
}

{\color{black} \section{Robustness of the Mean-field approximation of the networked eco-evolutionary dynamics}}

\begin{figure*}[h]
		\centering
\includegraphics[width=0.9\linewidth]{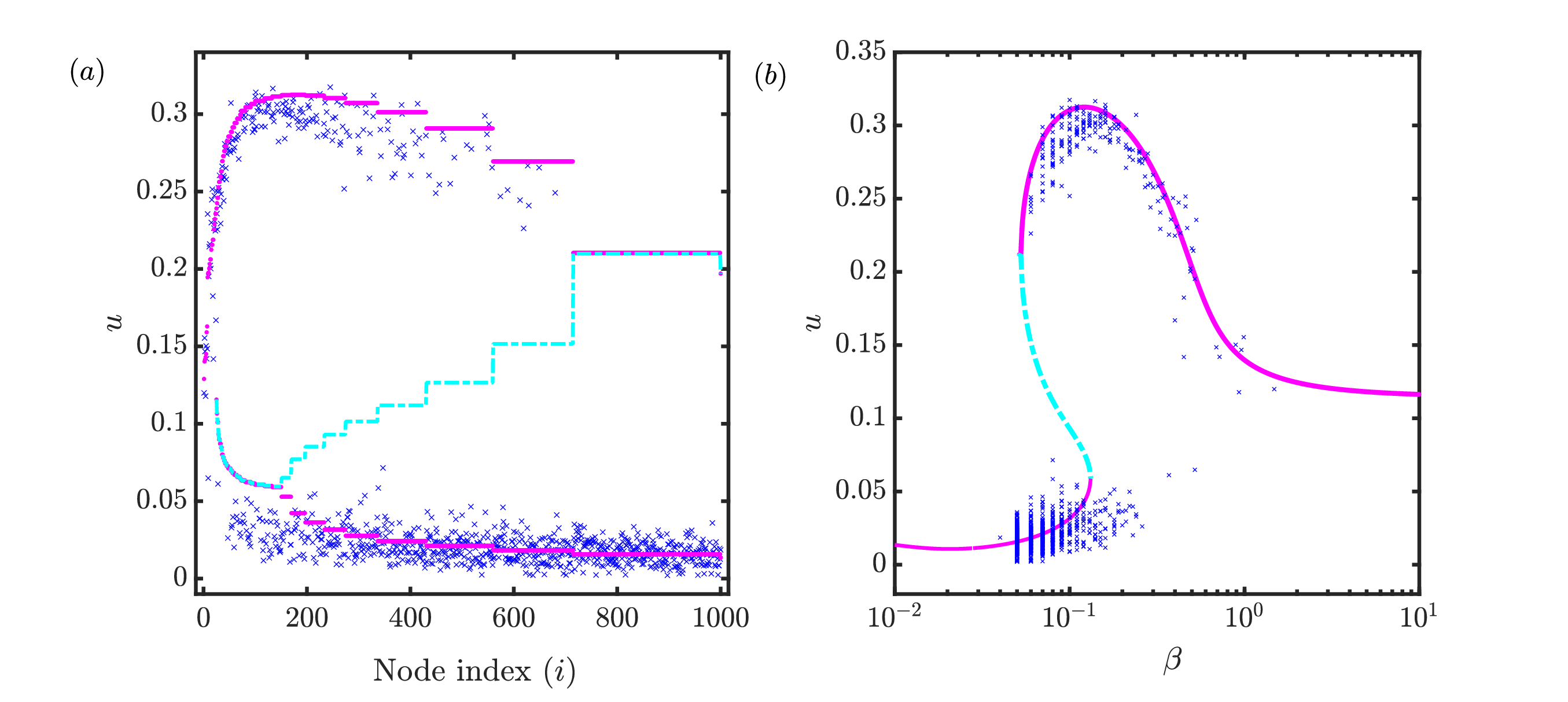} 
	\caption{\color{black}
    \textbf{Degree-Based Mean-Field Approximation and Bifurcation Structure:}
($a$) Degree-resolved stationary cooperator densities $u_i$ plotted against node index $i$ (nodes ordered by decreasing degree) for $\kappa = 40$ and $\epsilon=0.01$. Blue crosses denote the cooperator densities obtained from full network simulations, while the magenta (cyan) curves represent the corresponding stationary stable (unstable) solutions predicted by the reduced degree-based mean-field equation. ($b$) Bifurcation diagram of the reduced mean-field system as a function of the effective coupling parameter $\beta=\epsilon k$ for $\epsilon=0.01$. Magenta (cyan) curves denote the stable (unstable) stationary branches of cooperators, while blue crosses correspond to the cooperator densities obtained from the network simulations for $\kappa=40$. The agreement between the numerical data and the mean-field branches shows that node degree controls the transition between low- and high-cooperation states in the network.
}
\label{sf_kp_40}
\end{figure*}

{\color{black}
\subsection{Mean-field validation on scale-free networks for reduced diffusion asymmetry ($\kappa=40$)}

To further examine the mechanism underlying the emergence of cooperation-prone nodes in a complex network, we apply again the degree-based mean-field validation discussed in the
main text for a reduced diffusion rate ratio $\kappa=40$. All other
parameters and network properties remain identical to those used in the
main simulations, namely, a Barabási–Albert scale-free network with
$M=1000$ nodes and average degree $\langle k\rangle=10$. The purpose of
this analysis is to understand how decreasing the diffusion ratio
affects the bifurcation structure predicted by the reduced mean-field system 
\begin{equation*}
\begin{array} {lcl}
\dot u &= u\big[(1-u-v)(f_C(u,v)+b)-d\big] 
         + \beta (\langle u \rangle - u), \\
\dot v &= v\big[(1-u-v)(f_D(u,v)+b)-d\big] 
         + \kappa \beta (\langle v \rangle - v), \label{eq:mf_v}
\end{array}
\end{equation*} and the corresponding node-level organization observed in the full network dynamics.

Fig. ~\ref{sf_kp_40}$(a)$ compares the stationary cooperator densities
obtained from direct simulations of the full network system Eq. (3) (blue crosses in Fig.~\ref{sf_kp_40}$(a)$) of the main text with the stationary branches predicted by the
degree-based mean-field approximation. The blue crosses represent the
node-wise stationary values of $u_i$, while the solid and dash-dotted curves in Fig. ~\ref{sf_kp_40}$(a)$ denote
the stationary branches of the reduced system evaluated at
$\beta_i=\epsilon k_i$. The magenta curves correspond to stable
solutions of the reduced equations, whereas the cyan dash-dotted curve
represents the unstable branch separating the two stable states.

Even for the reduced diffusion ratio $\kappa=40$, the bifurcation
structure predicted by the mean-field system remains clearly visible.
Low-degree nodes cluster near the lower branch characterized by very
small cooperator densities, indicating that peripheral nodes remain dominated by the
defectors. In contrast, nodes with sufficiently large degree
shift toward the upper branch, where the stationary cooperator density
is substantially higher. The gradual transition between these two
branches produces the heterogeneous organization observed across the
network.

Further insight is obtained from Fig.~\ref{sf_kp_40}$(b)$, which shows
the bifurcation structure of the reduced mean-field system as a function
of the effective coupling parameter $\beta=\epsilon k$.  The diagram exhibits a saddle-node structure with two stable branches separated by an unstable one. For small values of $\beta$, corresponding to nodes with weak diffusive
coupling, the system remains close to the defector-dominated state (lower cooperative branch). As $\beta$ increases, a bistable region emerges in which two stable solutions coexist. Beyond the saddle-node threshold, the system evolves toward the upper cooperative branch.

Because the effective coupling parameter $\beta$ scales linearly with
node degree, this bifurcation structure directly translates into the
degree-dependent organization of the network dynamics. Highly connected nodes experience stronger effective coupling and therefore are more likely to cross the saddle-node threshold and settle on the cooperative branch, whereas peripheral nodes remain trapped on the lower defector-dominated branch.

Comparing these results with the stronger asymmetry case
$\kappa=100$ presented in Fig. 5 of the main text
reveals an important effect of diffusion asymmetry. When the diffusion ratio is reduced from $\kappa=100$ to $\kappa=40$, the qualitative bifurcation structure remains unchanged, but the separation between the stable branches becomes less pronounced. Consequently, the fraction of nodes occupying the cooperative branch decreases and the spread of simulation points around the theoretical curves becomes larger. This indicates that weaker mobility asymmetry reduces the strength of the effective coupling mechanism that drives the formation of cooperative hubs.

Overall, Fig.~\ref{sf_kp_40} confirms that the degree-based mean-field approximation continues to accurately capture the node-level dynamics even when the diffusion asymmetry is reduced. The heterogeneous cooperation patterns observed in the full simulations therefore arise from the same underlying bifurcation mechanism, although their magnitude and clarity depend sensitively on the diffusion ratio $\kappa$.
}

{\color{black}
\subsection{Mean-field validation on other network topologies}

To assess whether the mechanism underlying the emergence of zones dominated by cooperators depends on the specific structure of scale-free networks, we perform numerical experiments on alternative network topologies. In particular, we consider Erdős–Rényi random and small-world networks while keeping all model parameters identical to those used in the main text. The diffusion asymmetry is fixed at $\kappa = 100$, corresponding to the strong mobility imbalance regime analyzed in Fig. 5 of the main text. The goal of this analysis is to determine whether the degree-based mean-field approximation derived, continues to explain the observed
node-level dynamics in networks with different connectivity structures.

\subsubsection{Erdős–Rényi random network}
\begin{figure*}[h]
		\centering
\includegraphics[width=0.9\linewidth]{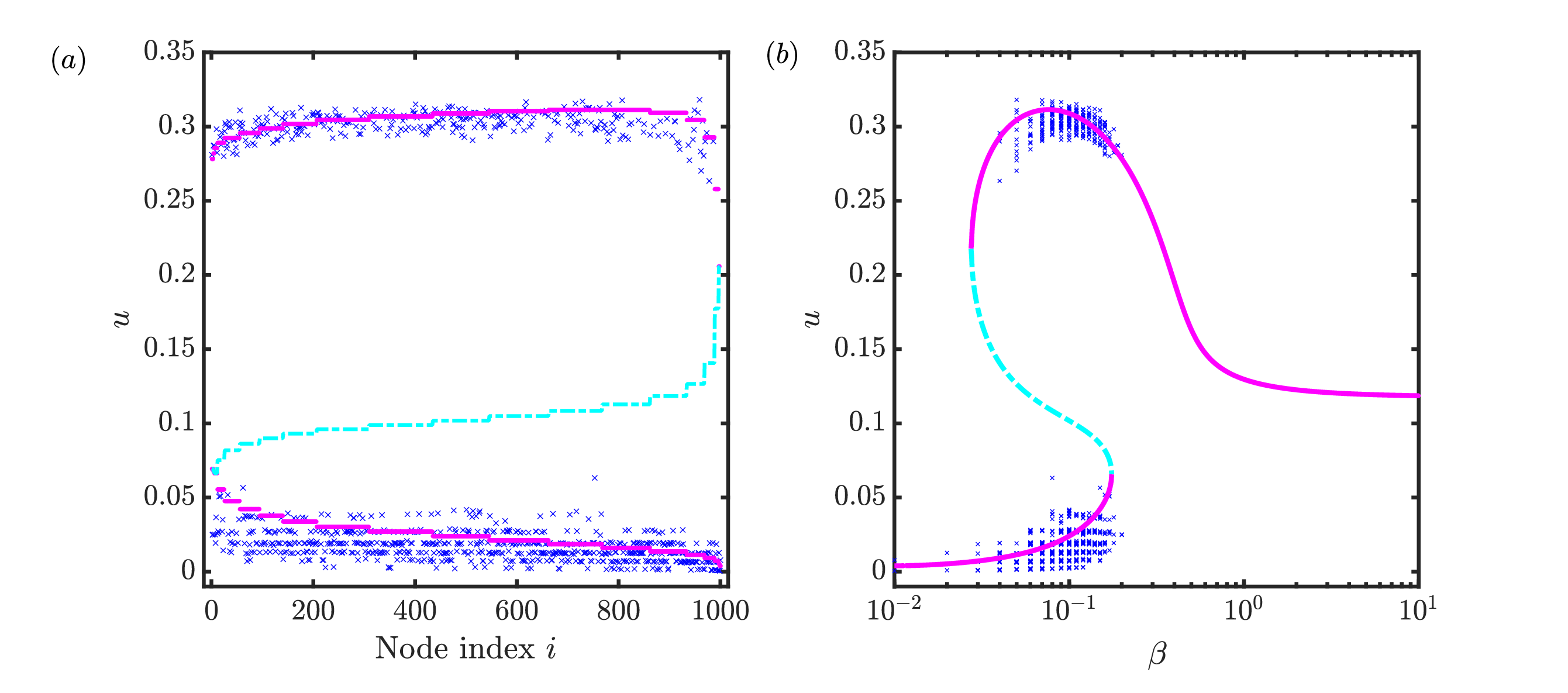} 
	\caption{\color{black}
\textbf{Mean-field validation of node-level dynamics on an Erdős–Rényi random network ($\kappa=100$).}
$(a)$ compares the stationary cooperator densities obtained from full network simulations with the predictions of the reduced mean-field equations for an Erdős–Rényi network with $M=1000$ nodes and average degree $\langle k\rangle=10$. Blue crosses denote the stationary values $u_i$ from numerical integration of system, while the magenta curves correspond to stable branches of the reduced system evaluated at $\beta_i=\epsilon k_i$. The cyan dashed curve represents the unstable branch separating the two stable states. Nodes are ordered by increasing degree. Due to the relatively narrow Poisson degree distribution, most nodes occupy a limited region of the theoretical branches. 
$(b)$ shows the bifurcation diagram of the reduced mean-field system as a function of the effective coupling parameter $\beta$. The characteristic saddle-node structure remains visible, and the simulation points align with the predicted stable branches, confirming that the degree-based mean-field approximation captures the node-level dynamics even in networks with weak structural heterogeneity.
}
\label{er_kp_100}
\end{figure*}

We first examine an Erdős–Rényi random network with $M=1000$ nodes and average degree $\langle k \rangle = 10$. In this topology, links are placed between node pairs with equal probability, leading to a binomial (approximately Poisson) degree distribution. Compared with scale-free networks, this structure exhibits significantly weaker degree heterogeneity, which directly influences the range of effective coupling strengths $\beta_i = \epsilon k_i$ experienced by individual nodes.

The comparison between the stationary node densities obtained from the full network simulations and the predictions of the reduced mean-field equations is shown in Fig.~\ref{er_kp_100}$(a)$. The blue crosses denote the stationary cooperator densities $u_i$ obtained from direct integration of the networked system, while the
colored curves correspond to the stationary branches predicted by the
mean-field equations evaluated at $\beta_i=\epsilon k_i$. The magenta
curves represent stable solution branches of the reduced system, and
the cyan dashed curve denotes the unstable branch separating the two
stable states.

Because the Erdős–Rényi network has a relatively narrow degree
distribution, most nodes experience similar effective coupling
strengths. Consequently, the simulation points cluster within a narrow region of the theoretical branches, producing a more homogeneous distribution of stationary densities than observed in scale-free networks. Nevertheless, the node-level densities remain well aligned with the predicted stable branches, indicating that the mean-field reduction continues to capture the essential dynamics of the full network system.

This interpretation is further supported by Fig.~\ref{er_kp_100}$(b)$, which shows the bifurcation structure of the reduced mean-field system as a function of the effective coupling parameter $\beta$. The diagram again exhibits the characteristic saddle-node bifurcation, with two stable branches separated by an unstable one. The simulation points occupy a relatively narrow band of $\beta$ values, reflecting the
limited variability in node degrees of the Erdős–Rényi topology.

\subsubsection{Small-world network}
\begin{figure*}[h]
		\centering
\includegraphics[width=0.9\linewidth]{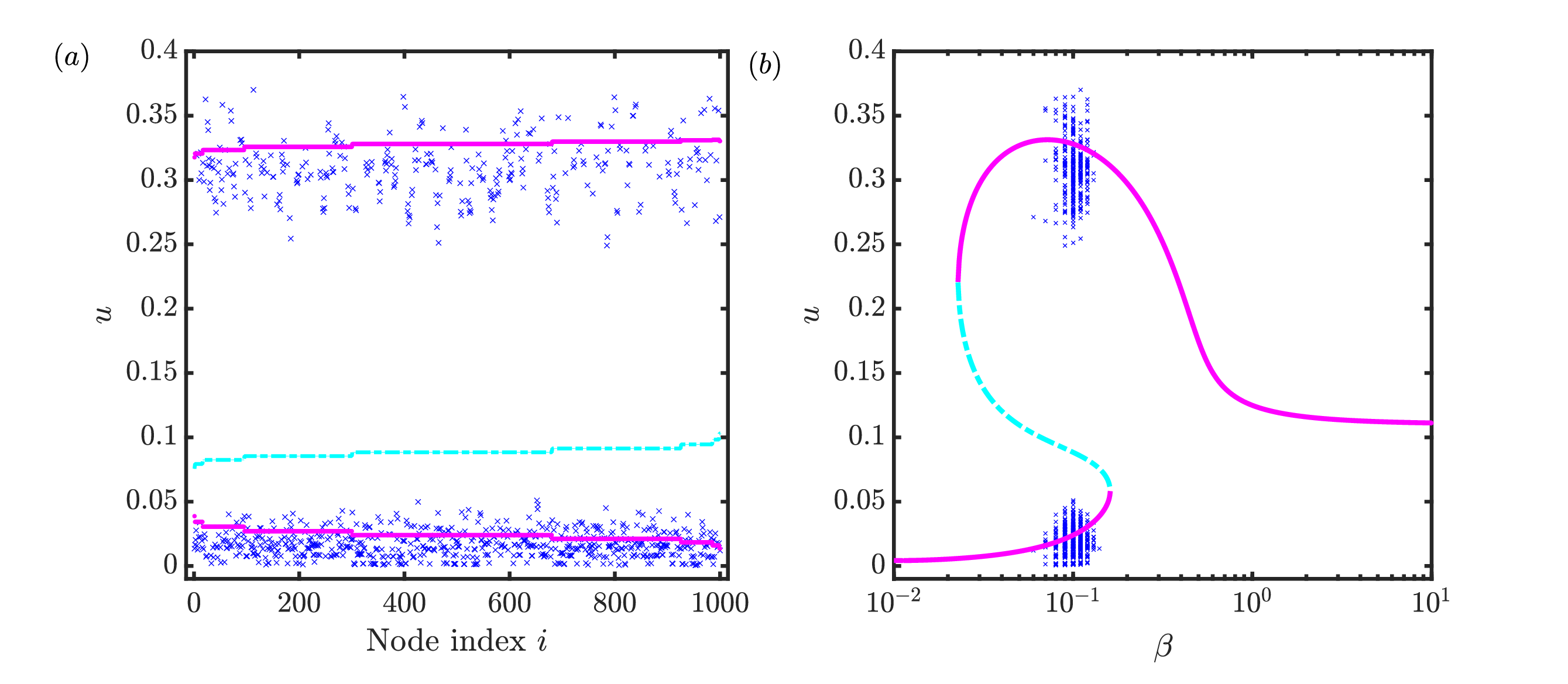} 
	\caption{\color{black}
\textbf{Mean-field validation of node-level dynamics on a small-world network ($\kappa=100$).}
$(a)$ shows the stationary cooperator densities $u_i$ obtained from simulations of the full network system on a Watts–Strogatz small-world network with $M=1000$ nodes, average degree $k=10$, and rewiring probability $p_{sw}=0.15$. Blue crosses represent the simulation results, while the magenta curves denote the stable stationary branches predicted by the reduced mean-field equations. The cyan dashed curve indicates the unstable branch separating the two stable solutions. Nodes are ordered by increasing degree. Because the small-world topology has a narrow degree distribution, most nodes experience similar effective coupling $\beta_i=\epsilon k_i$, resulting in a relatively concentrated distribution of stationary densities. 
$(b)$ shows the bifurcation structure of the reduced mean-field system as a function of the effective coupling parameter $\beta$. The diagram exhibits a saddle-node bifurcation with two stable branches separated by an unstable one. The clustering of simulation points within a narrow $\beta$ interval reflects the limited variability of node degrees in the small-world topology.
}
\label{sw_kp_100}
\end{figure*}
We next consider a Watts–Strogatz small-world network, generated with
$M=1000$ nodes, average degree $k=10$, and rewiring probability
$p_{sw}=0.15$. This topology preserves a relatively narrow degree
distribution while introducing shortcuts that reduce the average path
length of the network. Consequently, compared with scale-free networks, the heterogeneity in node degree is significantly reduced.

The comparison between the stationary cooperator densities obtained from the full network simulations and the predictions of the reduced
mean-field equations is shown in Fig.~\ref{sw_kp_100}$(a)$. The blue
crosses represent the stationary values of $u_i$ obtained from direct
integration of the networked system Eq. (3) of our main piece of study, while the colored curves correspond to the stationary branches predicted by the mean-field equations evaluated at $\beta_i=\epsilon k_i$. The magenta curves denote stable branches of the reduced system, whereas the cyan dash-dotted curve corresponds to the unstable branch separating the two stable states.

Because the degree distribution of the small-world network is relatively narrow, most nodes experience similar effective coupling strengths $\beta_i=\epsilon k_i$. As a result, the simulation points cluster around a limited region of the bifurcation diagram and the separation between distinct dynamical regimes becomes less pronounced than in the scale-free case. Nevertheless, the node densities remain consistent with the stable branches predicted by the mean-field approximation, confirming that the reduced equations still capture the underlying dynamics.

This behavior is further illustrated in Fig.~\ref{sw_kp_100}$(b)$, which shows the bifurcation structure of the reduced mean-field system as a function of the effective coupling parameter $\beta$. The diagram again exhibits a saddle-node structure with two stable branches separated by an unstable one. However, because the small-world network lacks the broad degree heterogeneity characteristic of scale-free networks, most nodes occupy a narrow range of $\beta$ values. Consequently, the coexistence of cooperative and defector-dominated nodes becomes less pronounced and the system tends to concentrate near the cooperative branch of the bifurcation diagram.

Overall, these results indicate that the mechanism responsible for cooperative hubs is not restricted to scale-free networks. Instead, the degree-dependent effective coupling predicted by the mean-field
approximation remains valid across different network topologies.
However, the extent of spatial differentiation depends strongly on the degree heterogeneity of the underlying network.

}

\section{Description of the supplementary movies}
\begin{itemize}
    \item \textbf{Movie S1:}This movie illustrates the breakdown of a stable coexistence between cooperators and defectors, represented by blue and red solid dots, respectively, in a scale-free (Barabási–Albert) network consisting of 1000 nodes with an average degree of $\langle k \rangle = 10$; this is the configuration considered throughout our study. The video captures the gradual emergence and evolution of spatio-temporal patterns in the network, depicting how the initially stable abundance of strategic individuals at each node destabilizes over time (from $t = 0$ to $t = 1000$) as a result of the introduced mobility effect. Nodes are ordered from highest to lowest degree.
    \item \textbf{Movie S2:} This movie depicts the disruption of the stable coexistence between cooperators and defectors, represented by blue and red solid dots, respectively, in a Watts–Strogatz network of $1000$ nodes with an average degree $\langle k \rangle = 10$ and a rewiring probability of ($p_{sw} = 0.15$). The video illustrates the gradual emergence of spatio-temporal structures within the network, where the initial homogeneous coexistence of strategic individuals destabilizes over time $t=0$ to $t = 1000$, due to the influence of mobility, leading to complex and dynamic spatial pattern formation. Nodes are ordered from highest to lowest degree.
    \item \textbf{Movie S3:} This movie presents the transition from a stable coexistence of cooperators and defectors, shown by blue and red solid dots, respectively, to a dynamically varying spatial configuration in an Erdős–Rényi random network of $1000$ nodes with an average degree $\langle k \rangle = 10$. The video captures how the introduction of mobility disrupts the initially stable distribution of strategies across the network, leading to the spontaneous development of irregular spatio-temporal patterns as time progresses from $t = 0$ to $t = 1000$. Nodes are ordered from highest to lowest degree.
\end{itemize}

\end{document}